\documentclass[conference]{IEEEtran}
\IEEEoverridecommandlockouts
\usepackage{cite}
\usepackage{amsmath,amssymb,amsfonts}
\usepackage{algorithmic}
\usepackage{graphicx}
\usepackage{textcomp}
\usepackage{xcolor}
\usepackage{hyperref}
\hypersetup{
    colorlinks=true,
    linkcolor=blue,
    filecolor=magenta,      
    urlcolor=cyan,
}

\usepackage{xcolor}
\usepackage{listings}
\usepackage{braket}
\usepackage{graphicx}
\usepackage{tikz}
\usepackage{xspace}
\usepackage{subcaption}
\usepackage{adjustbox}
\usetikzlibrary{quantikz}
\usetikzlibrary{automata, positioning, arrows}

\newcommand{\aswap}{\push{\bigotimes}}

\newcommand{\ifthen}[1]{\ensuremath{\mathit{if}~#1}}

\def\BibTeX{{\rm B\kern-.05em{\sc i\kern-.025em b}\kern-.08em
    T\kern-.1667em\lower.7ex\hbox{E}\kern-.125emX}}
\begin{document}

\pagenumbering{arabic}

\title{Relaxed Peephole Optimization: A Novel Compiler Optimization for Quantum Circuits}

\author{\IEEEauthorblockN{Ji Liu}
\IEEEauthorblockA{\textit{North Carolina State University} \\
Raleigh, USA \\
\texttt{jliu45@ncsu.edu}}
\and
\IEEEauthorblockN{Luciano Bello}
\IEEEauthorblockA{\textit{IBM Research} \\
Yorktown Heights, USA \\
\texttt{luciano.bello@ibm.com}}
\and
\IEEEauthorblockN{Huiyang Zhou}
\IEEEauthorblockA{\textit{North Carolina State University} \\
Raleigh, USA \\
\texttt{hzhou@ncsu.edu}}
}

\maketitle

\begin{abstract}
As in classical computing, compilers play an important role in quantum computing. Quantum processors typically support a limited set of primitive operations or quantum gates and have certain hardware-related limitations. A quantum compiler is responsible for adapting a quantum program to these constraint environments and decomposing quantum gates into a sequence of the primitive ones. During the compilation process, it is also critical for the compiler to optimize the quantum circuits in order to reduce the noise in the computation results. Since the noise is introduced by operations and decoherence, reducing the gate count is the key for improving performance.

In this paper, we propose a novel quantum compiler optimization, named relaxed peephole optimization (RPO) for quantum computers. RPO leverages the single-qubit state information that can be determined statically by the compiler. We define that a qubit is in a basis state when, at a given point in time, its state is either in the X-, Y-, or Z-basis ($\ket{+}/\ket{-}$, $\ket{L}/\ket{R}$ and $\ket{0}/\ket{1}$). When basis qubits are used as inputs to quantum gates, there exist opportunities for strength reduction, which replaces quantum operations with equivalent but less expensive ones. Compared to the existing peephole optimization for quantum programs, the difference is that our proposed optimization does not require an identical unitary matrix, thereby named `relaxed' peephole optimization. We also extend our approach to optimize the quantum gates when some input qubits are in known pure states. Both optimizations, namely the \textit{Quantum Basis-state Optimization (QBO)} and the \textit{Quantum Pure-state Optimization (QPO)}, are implemented in the IBM's Qiskit transpiler. Our experimental results show that our proposed optimization pass is fast and effective. The circuits optimized with our compiler optimizations obtain up to $18.0\%$ ($11.7\%$ on average) fewer $\mathit{CNOT}$ gates and up to $8.2\%$ ($7.1\%$ on average) lower transpilation time than that of the most aggressive optimization level in the Qiskit compiler. When running on real quantum computers, the success rates of 3-qubit quantum phase estimation algorithm improve by $2.30X$ due to the reduced gate counts.
\end{abstract}


\section{Introduction}
Quantum computing shows great potential in chemistry simulation~\cite{li2019variational},  combinatorial optimization~\cite{han2002combinatorial}, cryptography~\cite{mavroeidis2018impactcryptography}, machine learning~\cite{biamonte2017qml}, etc. Recently, Google, IBM, and Intel have announced their quantum computers with 72, 53, and 49 qubits, respectively~\cite{IBM53qubit,Intel49qubit,Google72qubit}. 
These noisy quantum computers are capable of running some quantum algorithms and would be helpful for exploiting the physics of many entangled particles~\cite{preskill2018NISQera}. However, state-of-the-art quantum computers do not have enough qubits to accommodate error correction codes, and the noise in the quantum computers hinders the development of quantum computing. 

As quantum computers typically support a limited set of basic operations/gates, quantum compilers/transpilers are responsible for decomposing complex quantum gates into the basic ones that the quantum computer supports. Quantum compilers also optimize quantum circuits to reduce the overall gate count or circuit depth. Since the accuracy of the final result can be affected by the system noise, such optimization is extremely important for quantum computers. 
The existing quantum compilers~\cite{qiskit,sivarajah2020tiket} exploit many optimization techniques. An important one is peephole optimization or operator strength reduction~\cite{sivarajah2020tiket}. The peephole optimization is analogous to its homonym in classical computing. The compiler traverses through the quantum circuit to find specific patterns of sub-circuits and substitute them with equivalent ones that have less primitive operations or shorter depth.
These substitutions keep the semantics of the quantum program as their unitary matrix representations are identical.

In this paper, we propose a new compiler optimization termed relaxed peephole optimization (RPO). 
It builds upon the fact that some of the qubit states for many quantum gates can be known or derived at compile-time.
This presents opportunities for 
replacing quantum operations with equivalent but less expensive ones. But the difference from the existing peephole optimization is that the unitary matrix of the circuit may change 
although the circuit functionality remains the same. For example, for a $\mathit{CNOT}$ gate, if the control qubit is in the $\ket{1}$ state, it is functionally equivalent to a $\mathit{NOT}$ gate on the target qubit. After determining the state of the control qubit, our compiler optimization will replace the $\mathit{CNOT}$ gate with a $\mathit{NOT}$ gate. However, the unitary matrices of $\mathit{CNOT}$ and $\mathit{NOT}$ gates are different and the peephole optimization would not be able to take advantage of such opportunities. In other words, our optimization can be viewed as a relaxed type of peephole optimization, which finds functionally equivalent circuits under certain circumstances. In our paper, we derive RPO for a wide range of quantum gates when some of their inputs are in known basis/pure states. 


In order to figure out the quantum states for RPO, we propose a quantum state analysis approach. With this analysis, we develop two optimization passes, namely the \textit{ Quantum Basis-state Optimization (QBO)} pass and the \textit{Quantum Pure-state Optimization (QPO)} pass. In our paper, we define $\ket{0}$, $\ket{1}$, $\ket{+}$, $\ket{-}$, $\ket{L}$, and $\ket{R}$ as basis states. The QBO pass identifies these basis states for every qubit during execution and optimizes quantum gates accordingly. The QPO pass determines single-qubit pure states and performs corresponding quantum circuit optimization. We also introduce annotations such that the programmer can guide the compiler optimization. We experimented with several quantum benchmarks on IBM Q quantum simulators and quantum computers. The experiments show that the circuits optimized using our approach have up to $18.0\%$ ($11.7\%$ on average) fewer $\mathit{CNOT}$ gates with up to $8.2\%$ ($7.1\%$ on average) less transpilation time than that of the most aggressive optimization level in the Qiskit compiler. Since other quantum compilers, e.g.,  $t\ket{ket}$~\cite{sivarajah2020tiket} use a similar gate model to IBM Qiskit, we expect that our proposed idea is applicable to them as well.

The major contributions of this work are listed as follows:
\begin{itemize}
\item We propose a new compiler optimization, relaxed peephole optimization (RPO).
\item We derive a comprehensive list of circuit optimizations for a wide range of quantum gates when some of their inputs are in known basis or pure states.
\item We present a quantum state analysis approach which identifies basis and pure states for each qubit in a quantum circuit. We also introduce annotations to enable users to provide information to facilitate state analysis.
\item We implement both QBO and QPO as compiler optimization passes in the IBM Qiskit transpiler.
\item We show that our proposed RPO achieves better results than the most aggressive optimization level in Qiskit. 
\end{itemize}

The remainder of the paper is organized as follows. Section~\ref{sec:background} introduces the background of quantum computing and quantum compilers. Sections~\ref{sec:Const} and~\ref{sec:fromConsttoPure} describe our findings on optimizing $\mathit{CNOT}$ and $\mathit{SWAP}$ gates with zero states and known pure states, respectively. Section~\ref{sec:generalization} generalizes the optimization for a broad range of quantum gates. Section~\ref{sec:quantumstateanalysis} discusses our compiler scheme to determine the quantum states of each qubit in a quantum circuit. Section~\ref{sec:methodology} presents our compiler implementation using IBM Qiskit. Our experimental results are discussed in Section~\ref{sec:performance}. Finally, Section~\ref{sec:conclusion} concludes the paper.

\section{Background and Related work}
\label{sec:background}
\subsection{Quantum Computing}
Qubit (quantum bit) is the basic unit of quantum information. Besides the classical states $\ket{0}$ and $\ket{1}$, a qubit can stay in any superposition state.
A superposition state can be represented as $\ket{\psi} = a\ket{0} + b\ket{1}$ where $a$ and $b$ are complex numbers and $\left | a \right |^2 + \left | b \right |^2= 1$. When measuring a superposition state $\ket{\psi}$ in the computational basis, the probability of getting $\ket{0}$ and $\ket{1}$ states are $\left | a \right |^2$ and $\left | b \right |^2$, respectively. The state of a quantum system is represented by a vector in a Hilbert space~\cite{nielsen2002quantumcomputation}: a complex vector space with an inner product. The state of multiple qubits can be expressed as the tensor product of each qubit state if they are not entangled: $\ket{\psi_{12}} = \ket{\psi_1}\otimes\ket{\psi_2} = \ket{\psi_1\psi_2}$. Entanglement is an unique property of quantum computing. When two qubits are entangled, their measurement results are correlated and the two-qubit state can not be expressed as the tensor product of individual qubits. For example, the two-qubit Bell state $\frac{1}{\sqrt{2}}(\ket{00} + \ket{11})$ is an entangled state. When the measurement outcome of the first qubit is $\ket{0}$, the measurement outcome of the second qubit must be $\ket{0}$.

A pure quantum state can be represented by a single state vector $\ket{\psi}$ in the Hilbert space. A mixed quantum state can not be represented in this way and corresponds to a probability mixture of pure states. A density matrix~\cite{blum2012density} $\rho = \sum_i P_i\ket{\psi_i}\bra{\psi_i}$ is used to describe the mixed states, where $P_i$ is the probability of the pure state $\ket{\psi_i}$. Generally speaking, a qubit system is in a pure state when the qubits in the system are not entangled with qubits out side the system. When the qubits are entangled with others, they are in the mixed state. For example, the two-qubit bell state is a two-qubit pure state since these two qubits are not entangled with others. The state can be represented by state vector $\frac{1}{\sqrt{2}}(\ket{00} + \ket{11})$. If we consider each qubit individually, however, since the two qubits are entangled, either qubit is in the mixed state with the density matrix as $\rho = \frac{1}{2}(\ket{0}\bra{0} + \ket{1}\bra{1})$. An n-qubit pure state $\ket{\psi}$ can be generated by applying an n-qubit unitary gate $U$ to n qubits, which are all in the $\ket{0}$ state: $\ket{\psi} = U\ket{0}^{\otimes n}$. The proof is in Appendix~\ref{appendix:purestate}. A key conclusion from the proof is that when a qubit is not entangled with the others, it is in a single-qubit pure state and this state can be generated by applying a single-qubit gate to a qubit in the $\ket{0}$ state. For simplification purposes, we will refer to single-qubit pure states as \emph{pure state} in the rest of this paper.

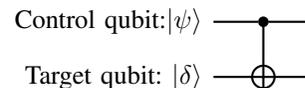
\begin{figure}[ht]
\centering
\begin{quantikz}
\lstick{Control qubit:$\ket{\psi}$} &\ctrl{1} & \qw\\
\lstick{Target qubit: $\ket{\delta}$} &\targ{} & \qw
\end{quantikz}
\caption{A $\mathit{CNOT}$ gate with its control and target qubit.}
\label{fig:quantumgates}
\end{figure}

A quantum program is essentially a sequence of instructions/gates operating on qubits. There are single-qubit gates such as Identity gate $I$ or $\mathit{id}$, Hadamard gate $\mathit{H}$, Phase changing gate $S$ and $T$, Pauli Gates $X$, $Y$, and $Z$, and two-qubit gates such as controlled-$\mathit{NOT}$ ($\mathit{CNOT}$) gate. The two qubits in the $\mathit{CNOT}$ gate are termed as control and target qubit as illustrated in Figure~\ref{fig:quantumgates}.
Instructions, unlike gates, are not necessarily reversible might include classical aspects in the quantum program. Example of such instructions include $\mathit{RESET}$, $\mathit{MEASURE}$, and conditional control. On this work only considers $\mathit{RESET}$ for simplicity.
The state-of-the-art quantum computers only support a set of basic gates/instructions. and the quantum compiler needs to decompose the quantum program into the supported primitives. For example, the IBM quantum computers support basis gates including four types of single-qubit gates $u1,u2,u3,id$, and a two-qubit $\mathit{CNOT}$ gate $cx$~\cite{IBMerrorrates}. 

\subsection{Qiskit Framework}




Qiskit~\cite{qiskit} is an open-source quantum computing software development framework. One element, named Qiskit Aqua, allows programmers to write codes for quantum algorithms.
Another element called Qiskit Terra provides a transpiler, which is responsible for quantum gate decomposition,  logical-to-physical qubit mapping, and circuit optimization. 
The transpiler consists of modular transpiler passes for circuit transformations. The transpiler pass manager schedules the transpiler passes and allows them to communicate. The users can control the pass manager to perform selective optimizations to the circuit. The transpiler has four pre-defined pass managers corresponding to the optimization level from 0 to 3. The higher optimization level, the higher the transpiler effort to optimize the circuit at the cost of transpilation time.

The different optimization levels can be described as follows~\cite{qiskit}:
Level 0 maps the circuit to the quantum device, with no explicit optimization included.
Level 1 maps the circuit and also performs light-weight optimizations such as collapsing adjacent gates.
Level 2 and 3 include noise-aware optimizations. Level 2 chooses noise-adaptive layout for the mapping and performs gate-cancellation procedure based on gate commutation relationships. Level 3 extends the passes from level 2 to include re-synthesis of two qubit blocks in the circuit as well as more iterations in the stochastic routing process.
The re-synthesis process is performed by the \texttt{Collect2qBlocks} and the \texttt{ConsolidateBlocks} passes in Qiskit. The \texttt{Collect2qBlocks} pass traverses the circuit and collects sequences of gates acting on two qubits. The \texttt{ConsolidateBlocks} pass calculates the unitary matrices of two-qubit blocks and re-synthesizes them to more optimized circuits. This transpiler pass is similar to the operator strength reduction or peephole optimization in classical computing. The difference between our proposed optimization and the \texttt{ConsolidateBlocks} pass is that we do not preserve the unitary matrix although we replace the circuits with the ones that have the same functionality.
\subsection{Related Work}
Prior works have been proposed to optimize quantum circuits at the gate level. Venturelli et al.~\cite{venturelli2019constrain} proposed an automated, architecture-aware software framework aided by constraint programming. Nam et al.~\cite{nam2018automated} developed automated optimization methods using phase polynomials. There are also noise-adaptive compilers~\cite{murali2019noiseadaptive,tannu2019notallqubits,murali2020crosstalk} which take advantage of the noise characteristics of the target backend to aid the optimization.

Circuit equivalence has been discussed in terms of quantum algorithms~\cite{viamontes2007checkingequivalence}, quantum compiler optimization~\cite{prasad2006quantumpeephole} and quantum compilation verification~\cite{amy2018verification}. Two circuits are considered as equivalent when their unitary matrix representations are identical~\cite{amy2019formalequivalence}. Garcia-Escartin et al.~\cite{garcia2011equivalentcircuits} proposed a list of equivalent rules for identifying equivalent circuits. In compiler optimization, the process of substituting sub-circuits with their equivalent ones is termed as peephole optimization. 

Peephole optimization~\cite{mckeeman1965peephole} identifies small sets of instructions and substitutes them with equivalent sets that have better performance. Prasad et al.~\cite{prasad2006quantumpeephole} proposed a quantum circuit optimization algorithm that relies on peephole optimization. Sivarajah et al.~\cite{sivarajah2020tiket} introduced a compiler named $t\ket{ket}$ for noisy quantum computers. The peephole optimization in $t\ket{ket}$ compiler traverses through the circuit to find long sequences of single-/two-qubit gates and replaces them with the circuits generated by Euler and KAK decomposition~\cite{kraus2001kak1}. The similar optimization exists in other quantum compilers. For example, the transpiler in Qiskit Terra~\cite{qiskit} contains optimization pass \texttt{Collect2qBlocks} and \texttt{ConsolidateBlocks}, which collaboratively identify and substitute two-qubit blocks with equivalent circuits. The Cirq~\cite{cirq} framework also provides similar optimization for adjacent single- and two-qubit gates.

Hoare logic~\cite{haner2018hoarelogic} has been used to optimize quantum circuits. The optimizer removes trivial operations based on postconditions of the subroutines and the triviality conditions. The postconditions can be derived from the hoare triples of quantum subroutines. While our approach shares a few common optimization cases with the prior work, our optimization is more generic as we include optimizations for the quantum gates that are not trivial, e.g., the one shown in Eq.~\ref{eq:SWAPtoU}. Besides generality, the quantum state analysis discussed in Section~\ref{sec:quantumstateanalysis} provides fine-grained analysis for quantum states and enables more quantum gate optimizations. Moreover, since the hoare logic pass requires a classical $Z3$ solver~\cite{de2008z3} to express the conditions, it significantly increases the transpilation time. In Section~\ref{sec:performance}, we show that our optimization pass is faster and more effective than the hoare logic pass.

\section{Zero State}
\label{sec:Const}

We use a basic optimization, which only leverages a $\ket{0}$ state, as a stepping stone for more generic optimizations (see Section \ref{sec:generalization}).
We also introduce some handful notations and concepts.

Consider the following situation: if a $\mathit{CNOT}$ gate with the control qubit being in the $\ket{0}$ state, the $\mathit{CNOT}$ gate has no effect and can be removed or replaced with a wire:

\begin{equation} \label{eq:initialObsCNOT}
\centering
\begin{quantikz}
\lstick{$\ket{\phi}$} &\targ{1}  & \rstick{$\ket{\phi}$}\qw \\
\lstick{$\ket{\psi}$}    &\ctrl{-1} & \rstick{$\ket{\psi}$}\qw
\end{quantikz}
=\begin{quantikz}
 \lstick{$\ket{\phi}$}& \rstick{$\ket{\phi}$} \qw \\[2mm]
\lstick{$\ket{\psi}$} & \rstick{$\ket{\psi}$} \qw
\end{quantikz}
\ifthen{\ket{\psi} = \ket{0}}
\end{equation}

The unitary matrices of the $\mathit{CNOT}$ gate and the idle wire are obviously different, and they are not considered as equivalent in the existing compilers.
However, they are functionally equivalent when the control qubit is in $\ket{0}$ state.

\begin{figure}
\centering
\begin{adjustbox}{width=0.9\linewidth}
\begin{quantikz}[column sep=0.5mm]
\lstick{$\ket{\phi}$} & \swap{1} & \rstick{$\ket{\psi}$}\qw \\
\lstick{$\ket{\psi}$} & \swap{-1} &\rstick{$\ket{\phi}$}\qw
\end{quantikz}
{=}
\begin{quantikz}[column sep=1mm]
\lstick{$\ket{\phi}$}\qw & \targ{1}  &  \ctrl{1} & \targ{1}  & \rstick{$\ket{\psi}$}\qw \\
\lstick{$\ket{\psi}$}\qw & \ctrl{-1} & \targ{-1} & \ctrl{-1} & \rstick{$\ket{\phi}$}\qw
\end{quantikz}
{=}
\begin{quantikz}[column sep=1mm]
\lstick{$\ket{\phi}$}\qw  & \ctrl{1}  &  \targ{1} & \ctrl{1} & \rstick{$\ket{\psi}$}\qw \\
\lstick{$\ket{\psi}$}\qw &  \targ{-1} &  \ctrl{-1} & \targ{-1} &\rstick{$\ket{\phi}$}\qw
\end{quantikz}
\end{adjustbox}
\caption{$3$-$\mathit{CNOT}$ decomposition for $\mathit{SWAP}$ gate }
\vspace*{-6mm}
\label{fig:SWAPdecomposition}
\end{figure}
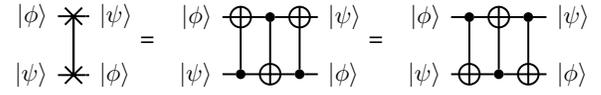

Although this special $\mathit{CNOT}$ example may seem trivial, we can leverage it for optimizing $\mathit{SWAP}$ gates.
$\mathit{SWAP}$ gates are often introduced 
during the logical-to-physical qubit mapping to allow quantum operations on qubits that not physically adjacent.
Being symmetric gates, $\mathit{SWAP}$ gates has two possible decompositions as showed in Figure~\ref{fig:SWAPdecomposition}.

Many times $\mathit{SWAP}$ gates are introduced in the beginning part of a circuit, 
when there is a good chance that one of the qubits to be swapped is still in the initial ground state $\ket{0}$.
In such cases, we can take advantage of a reduced $\mathit{SWAP}$ gate design, which only swaps any state $\ket{\psi}$ with the state $\ket{0}$:

\begin{equation} \label{eq:initialObsSWAP}
\begin{quantikz}[column sep=1mm]
\lstick{$\ket{\phi}$}\qw & \targ{1}  &  \ctrl{1} & \targ{1}  & \rstick{$\ket{\psi}$}\qw \\
\lstick{$\ket{\psi}$}\qw & \ctrl{-1} & \targ{-1} & \ctrl{-1} & \rstick{$\ket{\phi}$}\qw
\end{quantikz}
{=}
\begin{quantikz}[column sep=0.5mm]
\lstick{$\ket{\phi}$} &\ctrl{1} & \targ{}  & \rstick{$\ket{\psi}$} \qw\\
\lstick{$\ket{\psi}$} &\targ{} & \ctrl{-1} &  \rstick{$\ket{\phi}$} \qw
\end{quantikz}
{\ket{\psi} = \ket{0}}
\end{equation}

Let us introduce a notation for this special 2-$\mathit{CNOT}$-$\mathit{SWAP}$ gate. It is referred to as $\mathit{SWAPZ}$ (swap-zero) in the remaining of the paper:

\begin{equation} \label{eq:swapz}
 \begin{quantikz}
\qw & \swap{1} & \qw \\
\qw & \push{\bigotimes} & \qw
\end{quantikz}
\triangleq
\begin{quantikz}
\qw &\ctrl{1} & \targ{}  & \qw \\
\qw &\targ{} & \ctrl{-1} & \qw
\end{quantikz}
\end{equation}

In summary, a $\mathit{SWAP}$ gate that involves a $\ket{0}$  state can be replaced for a less expensive $\mathit{SWAPZ}$.
This equivalence follows from Equations~\ref{eq:initialObsSWAP} and \ref{eq:swapz}.
The first $\mathit{CNOT}$ gate from the $\mathit{SWAP}$ decomposition can be removed as its control qubit is in $\ket{0}$ state.
The resulting $\mathit{SWAPZ}$ gate consists of two $\mathit{CNOT}$ gates, one less compared to the generic $\mathit{SWAP}$ gate:

\begin{equation} \label{eq:swapzZero}
\begin{quantikz}[column sep=2mm]
\lstick{$\ket{\phi}$}\qw & \swap{1} & \rstick{$\ket{\psi}$}\qw \\
\lstick{$\ket{\psi}$}\qw & \swap{-1} & \rstick{$\ket{\phi}$}\qw
\end{quantikz}
{=}
\begin{quantikz}[column sep=2mm]
\lstick{$\ket{\phi}$}\qw & \swap{1} & \rstick{$\ket{\psi}$}\qw \\
\lstick{$\ket{\psi}$}\qw & \aswap & \rstick{$\ket{\phi}$}\qw
\end{quantikz}
\ifthen{\ket{\psi} = \ket{0}}
\end{equation}

\section{$\mathit{SWAP}$ on Pure States}
\label{sec:fromConsttoPure}
It is possible to extend the explained zero-state optimization of a $\mathit{SWAP}$ gate to any known single-qubit pure state.

A qubit is in a pure state up to a certain point if all the operations on that qubit up to that point do not entangle with other qubits.
A pure state can be represented by a single state vector $\ket{\pi}$ in the Hilbert space.
Based on the derivation in Appendix~\ref{appendix:purestate}, the single-qubit pure state $\ket{\pi}$ can be obtained by applying a single unitary gate $\mathit{U}$ to the $\ket{0}$ state, $\ket{\pi} = U\ket{0}$. Due to unitarity, we can apply the inverse gate $U^{-1}$ to transform the qubit in the pure state $\ket{\pi}$ back to the state $\ket{0}$.

\begin{figure*}[!t]
\begin{equation}
\label{eq:SWAPtoASWAP}
\begin{quantikz}[column sep=2mm]
\lstick{$\ket{\psi}$} & \swap{1} & \rstick{$\ket{\pi}$} \qw\\
\lstick{$\ket{\pi}$} & \swap{-1} &  \rstick{$\ket{\psi}$} \qw
\end{quantikz}
=
\begin{quantikz}[column sep=2mm]
\lstick{$\ket{\psi}$} & \qw &\qw & \swap{1} & \phase[label position=above]{\ket{0}}& \gate{U} &\rstick{$\ket{\pi}$} \qw\\
\lstick{$\ket{\pi}$} & \gate{U^{-1}} & \phase[label position=above]{\ket{0}}
& \swap{-1} &\qw & \qw &\rstick{$\ket{\psi}$} \qw
\end{quantikz}
{=}
\begin{quantikz}[column sep=2mm]
\lstick{$\ket{\psi}$} & \qw &\qw& \swap{1} & \phase[label position=above]{\ket{0}}& \gate{U} & \rstick{$\ket{\pi}$} \qw\\
\lstick{$\ket{\pi}$} & \gate{U^{-1}}& \phase[label position=above]{\ket{0}} &\push{\bigotimes} &\qw & \qw & \rstick{$\ket{\psi}$} \qw
\end{quantikz}
\ifthen{\ket{\pi} = U\ket{0}}
\end{equation}
\vspace*{-6mm}
\end{figure*}

When a $\mathit{SWAP}$ gate operates on two qubits $\ket{\psi}$ and $\ket{\pi}$ and one of them, e.g., $\ket{\pi}$, is in a known pure state generated with $U\ket{0}$, the $\mathit{SWAP}$ gate can be replaced with a $\mathit{SWAPZ}$ gate. As shown in Equation~\ref{eq:SWAPtoASWAP}, the inverse gate $U^{-1}$ transforms the pure state $\ket{\pi}$ to state $\ket{0}$ (and back to $\ket{\pi}$ after the $SWAP$).
Then, the $\mathit{SWAP}$ gate is to swap state $\ket{0}$ with state $\ket{\psi}$, thereby can be optimized.

Altogether, when a $\mathit{SWAP}$ gate operates on an arbitrary state $\ket{\psi}$ and a known pure state $\ket{\pi}$, we can optimize the $\mathit{SWAP}$ gate into a $\mathit{SWAPZ}$ gate along with two single-qubit gates. This reduces the number of $\mathit{CNOT}$ gates by one and introduces two extra single-qubit gates. In the noisy quantum systems, the gate error of a $\mathit{CNOT}$ gate is usually much higher than the accumulated error of two single-qubit gates. For example, on \texttt{ibmq\_16\_melbourne}, the single-qubit gate error rates range from $10^{-4}$ to $10^{-3}$ while the two-qubit $\mathit{CNOT}$ gate's error rate is $10^{-2}$~\cite{IBMerrorrates}. Moreover, 
the single-qubit gate $U$ and $U^{-1}$ may be canceled or merged during the subsequent compiler optimization passes where multiple single-qubit gates are fused into one single-qubit gate. In those cases, we essentially reduce the number of $\mathit{CNOT}$ gates by one and do not introduce any additional gates.

Consider the special case where both $\ket{\psi}$ and $\ket{\pi}$ are in known pure states. We can always find a single-qubit unitary gate $V$ to transform $\ket{\psi}$ state to $\ket{\pi}$ state, $\ket{\pi} = V\ket{\psi}$. Similarly, the inverse gate $V^{-1}$ transforms the $\ket{\psi}$ state to $\ket{\pi}$ state, as shown in Equation~\ref{eq:SWAPtoU}. Therefore, we can substitute the $\mathit{SWAP}$ gate with two single-qubit gates $V$ and $V^{-1}$ in this special case.

\begin{equation} \label{eq:SWAPtoU}
\begin{quantikz}[column sep=1mm]
\lstick{$\ket{\psi}$} & \swap{1} & \rstick{$\ket{\pi}$} \qw\\
\lstick{$\ket{\pi}$} & \swap{-1} &  \rstick{$\ket{\psi}$} \qw
\end{quantikz}
= \begin{quantikz}[column sep=1mm]
\lstick{$\ket{\psi}$} & \gate{V} & \rstick{$\ket{\pi}$} \qw\\
\lstick{$\ket{\pi}$} & \gate{V^{-1}} &  \rstick{$\ket{\psi}$} \qw
\end{quantikz}
\ifthen{\ket{\pi} = V\ket{\psi}} 
\end{equation}

\section{Basis-State and Pure-State Optimization}
\label{sec:generalization}

Sections \ref{sec:Const} and \ref{sec:fromConsttoPure} showcase that for $\mathit{CNOT}$ and $\mathit{SWAP}$ gates, if some input qubit states can be determined, the gates can be simplified.
In this section, we generalize this optimization to more basis and pure states in the context of different types of quantum gates.
We term such optimizations as quantum basis-state optimization (QBO) and quantum pure-state optimization (QPO), respectively. 


\subsection{Optimizing Single-Qubit Gates}\label{sec:optSingleQubit}

For single-qubit gates, if the input is in a particular basis state, they can be simplified.
In particular, if the input state is an eigenstate of the gate and the eigenvalue is 1, i.e., $\ket{\phi} = U\ket{\phi}$ the gate U can be removed, as shown in Equation~\ref{eq:single-optimization}.
For example, an X gate with the input $\ket{+}$ has not effect and can be removed since $\ket{+} = X\ket{+}$.
\begin{equation}\label{eq:single-optimization}
\begin{quantikz}[column sep=2mm]
\lstick{$\ket{\phi}$} &\gate{U} & \rstick{$\ket{\phi}$}\qw
\end{quantikz}
=\begin{quantikz}[column sep=5mm]
\lstick{$\ket{\phi}$} & \rstick{$\ket{\phi}, \ifthen{\ket{\phi} = U\ket{\phi}}$}\qw
\end{quantikz}
\end{equation}

Realistically speaking, the opportunity for such single-qubit gate optimization is limited in useful quantum programs.
However, it lays the foundation for QBO for multi-qubit gates.

\newcommand{\cCX}{
\begin{tikzcd}[column sep=3mm]
\qw{} & \ctrl{1}  & \qw{} \\
\qw{} & \targ{-1} & \qw{}
\end{tikzcd}
}

\newcommand{\cI}{
\begin{tikzcd}[column sep=3mm]
\qw & \qw & \qw \\[2mm] 
\qw & \qw & \qw         
\end{tikzcd}
}

\newcommand{\cX}{
\begin{tikzcd}[column sep=3mm]
\qw & \qw & \qw \\[-0.5mm]  
\qw & \gate{X} & \qw        
\end{tikzcd}
}

\newcommand{\cZ}{
\begin{tikzcd}[column sep=3mm]
\qw & \gate{Z} & \qw \\ 
\qw & \qw & \qw         
\end{tikzcd} 
}

\begin{table}[h]
\centering
\caption[]{Equivalences for basis-states in 
\begin{adjustbox}{height=0.9\baselineskip}
\begin{quantikz}[row sep=3mm, column sep=3mm]
\lstick{$\ket{\phi}$}&\ctrl{1} & \qw \\
\lstick{$\ket{\psi}$} &\targ{-1} & \qw
\end{quantikz}
\end{adjustbox}
}
\label{tbl:closedCNOT}
\begin{adjustbox}{width=\linewidth}
\begin{math}
\begin{array}{r|c|c|c|c|c}
\tikz{\node[below left, inner sep=1pt] (target) {$\ket{\psi}$};%
      \node[above right,inner sep=1pt] (control) {$\ket{\phi}$};%
      \draw (target.north west|-control.north west) -- (target.south east-|control.south east);}
& 
\multicolumn{1}{c}{\top} & 
\multicolumn{1}{c}{\ket{0}} &
\multicolumn{1}{c}{\ket{1}} &
\multicolumn{1}{c}{\ket{+}} & 
\multicolumn{1}{c}{\ket{-}} \\
\cline{1-6} 
\top & \cCX & \cI & \cX & \cCX & \cCX \\
\cline{2-6}  
\ket{+} & \cI  & \cI & \cI & \cI & \cI \\
\cline{2-6} 
\ket{-} & \cZ & \cI & \cI & \cZ & \cZ \\
\cline{2-6} 
\ket{0} & \cCX & \cI & \cX & \cCX & \cCX \\
\cline{2-6} 
\ket{1} & \cCX & \cI & \cX & \cCX & \cCX \\
\end{array}
\end{math}
\end{adjustbox}
\end{table}

\subsection{Optimizing $\mathit{CNOT}$}
\label{subsec:constantopt}
Since the $\mathit{CNOT}$ gate is the essential component to entangle qubits, we first focus on the basis states that have special effect on $\mathit{CNOT}$ gates.
For $\mathit{CNOT}$ gates, the observation from Equation \ref{eq:initialObsCNOT} can be generalized to other basis besides $\ket{0}$. These basis states include $\ket{0}$, $\ket{1}$, $\ket{+}$, and $\ket{-}$.
The $\mathit{CNOT}$ gates can be optimized when their control qubits are in $\ket{0}$ or $\ket{1}$ state or when their target qubits are in $\ket{+}$ or $\ket{-}$ state. We list all the possible combinations and the corresponding optimized circuits in Table \ref{tbl:closedCNOT}, where $\top$ represents a state, which is not the X- or Z- basis ($\ket{+}$, $\ket{-}$, $\ket{0}$, $\ket{1}$).


When the target qubit is in the $\ket{+}$ state, i.e.,  $\frac{\ket{0} + \ket{1}}{\sqrt{2}}$, the $\mathit{X}$ gate has no effect as $\ket{+}$ is its eigenstate with eigenvalue of 1, i.e., $\ket{+} = X\ket{+}$. Therefore, the $\mathit{CNOT}$ gate can be replaced with wires no matter what state the control qubit is in. On the other hand, if the target qubit is in the state $\ket{-}$, i.e., $\frac{\ket{0} - \ket{1}}{\sqrt{2}}$, the $\mathit{CNOT}$ gate may be replaced with a wire on the target qubit and a $\mathit{Z}$ gate on the control bit.
The derivation is in Appendix~\ref{appendix:CNOTminus}. 

Furthermore, the $\mathit{Z}$ gate can be eliminated when the control qubit is known to be in $\ket{0}$ state.
Since $\ket{0}$ state is the eigenstate of Pauli matrix Z with eigenvalue equals to 1, applying the $\mathit{Z}$ gate will not change the state: $Z\ket{0} = \ket{0}$.

Similarly, we derive the optimized $\mathit{SWAP}$ gate for X- and Z-bases input combinations, which are shown in Appendix~\ref{appendix:swap}.
This optimization can be seen as a particular case of QPO, since a basis-state is also a pure-state.
However, implementing QBO separately is more efficient, since it requires fewer changes in the circuit. We also derive similar simplifications to  Table~\ref{tbl:closedCNOT} for the controlled-$\mathit{Z}$ gates with inputs in the Z-basis states ($\ket{0}$, $\ket{1}$).

\subsection{Optimizing Multi-Qubit Gates}
Besides $\mathit{CNOT}$ gates and $\mathit{SWAP}$ gates, QBO and QPO can be generalized to optimize a broader range of quantum gates.

First, QBO for $\mathit{CNOT}$ gates can be generalized to optimize the multi-controlled $\mathit{NOT}$ gates. The optimization of multi-controlled $\mathit{NOT}$ gates with some known inputs can be derived as follows. 1) If any of the control qubits is in the $\ket{0}$ state, we can remove the multi-controlled $\mathit{NOT}$ gate. 2) If any of the control qubits is in the $\ket{1}$ state, we can remove the control qubit and substitute the gate using a multi-controlled $\mathit{NOT}$ gate with one less control qubit. 3) If the target qubit is in $\ket{+}$ state, we can remove the multi-controlled $\mathit{NOT}$ gate. 4) If the target qubit is in $\ket{-}$ state, we can substitute the gate with a multi-controlled $\mathit{Z}$ gate with the target on any of the previous control qubits. In our study, we find that Toffoli gates are widely used in quantum algorithms. A Toffoli gate is a $\mathit{NOT}$ gate with two controlled qubits. Equation~\ref{eq:toffoli_optimization} presents the optimizations for the Toffoli gate. The open controlled $\mathit{NOT}$ gates can also be optimized using a similar approach as discussed is in Appendix~\ref{appendix:ClosedandOpen}.

Second, the optimization of multi-controlled $\mathit{NOT}$ gates can be further generalized to multi-controlled unitary gates. When the control qubits are in $\ket{0}$ or $\ket{1}$, the optimization rules are the same. Assume the unitary gate $U$ has eigenstates $\ket{\psi_+}$ and $\ket{\psi_-}$ with eigenvalues of 1 and -1, respectively. We have $\ket{\psi_+} = U\ket{\psi_+}$ and $\ket{\psi_-} = -U\ket{\psi_-}$. When the target qubit is in $\ket{\psi_+}$ and $\ket{\psi_-}$ the optimization rules are the same as the rules for multi-controlled $\mathit{NOT}$ gates with respect to $\ket{+}$ and $\ket{-}$ states.
\vspace*{-3mm}
\begin{equation} \label{eq:toffoli_optimization}
\vspace*{-3mm}
    \begin{tikzcd}[column sep=2mm]
    \lstick{$\ket{\pi}$} &\ctrl{1}  & \rstick{$\ket{\pi'}$}\qw \\
    \lstick{$\ket{\phi}$} &\ctrl{1} & \rstick{$\ket{\phi'}$}\qw \\
    \lstick{$\ket{\psi}$} &\targ{} & \rstick{$\ket{\psi'}$}\qw
    \end{tikzcd}
=
\begin{cases}
    \begin{tikzcd}[row sep=3mm]
     \qw & \qw &[4.5mm] \rstick[wires=3]{\textit{if} $\ket{\pi} = \ket{0}$\\[1mm]\textit{or} $\ket{\psi} = \ket{+}$} \qw \\
     \qw & \ghost{X}\qw &  \qw \\
     \qw & \qw &  \qw
     \end{tikzcd}\\
     \begin{tikzcd}[row sep=3mm]
     \qw &[1mm] \ghost{X}\qw & \rstick[wires=3]{\textit{if} $\ket{\pi} = \ket{1}$} \qw \\
     \qw & \ctrl{1} & \qw  \\
     \qw & \targ{}  & \qw 
     \end{tikzcd}\\
    \begin{tikzcd}[row sep=3mm]
     \qw & \ctrl{1} & \rstick[wires=3]{\textit{if} $\ket{\psi} = \ket{-}$} \qw \\
     \qw & \gate{Z} & \qw  \\
     \qw & \ghost{X}\qw  & \qw 
     \end{tikzcd}
\end{cases}
\end{equation}

Third, QBO/QPO for $\mathit{SWAP}$ gates can also be extended to the Fredkin gate (aka $\mathit{CSWAP}$ gate) and multi-controlled $\mathit{SWAP}$ gates. A Fredkin gate is a controlled $\mathit{SWAP}$ gate with a single control qubit. A Fredkin gate can be decomposed into two $\mathit{CNOT}$ gates and a Toffoli gate. The decomposition is included in Figure~\ref{fig:fredkin} in Appendix~\ref{appendix:fredkin}. In the same way, multi-controlled $\mathit{SWAP}$ gates can be decomposed into two $\mathit{CNOT}$ gates and a multi-controlled $\mathit{NOT}$ gate. If the control qubit $\ket{\psi}$ is in the $\ket{0}$ state, we can remove the Fredkin gate. If it is in the $\ket{1}$ state, we can substitute the Fredkin gate with a $\mathit{SWAP}$ gate. If any of the target states $\ket{\psi}$ or $\ket{\pi}$ is in a known basis state, we can optimize the first $\mathit{CNOT}$ gate accordingly. If both of the target states $\ket{\psi}$ and $\ket{\pi}$ are in known pure states, following the optimization in Equation~\ref{eq:SWAPtoU}, we can substitute the Fredkin gate with two controlled $U$ gates, as shown in Equation~\ref{eq:fredkintoU}, where gates $U$ and $U^{-1}$ have the relationship $\ket{\pi} = U\ket{\psi}$ and $\ket{\psi} = U^{-1}\ket{\pi}$. Since a Toffoli gate can be implemented with six $\mathit{CNOT}$ gates and eight single-qubit gates~\cite{shende2008cnottoffoli}, the Fredkin gate would need eight $\mathit{CNOT}$ gates and eight single-qubit gates following the decomposition in Figure~\ref{fig:fredkin}. In comparison, a controlled-U gate can be implemented with at most two $\mathit{CNOT}$ gates and four single-qubit gates~\cite{song2002optimalcontrolledU}. Therefore, our optimized Fredkin gate in Equation~\ref{eq:fredkintoU} would require at most four $\mathit{CNOT}$ gates and eight single-qubit gates. As a result, our proposed QPO reduces at least four $\mathit{CNOT}$ gates for a Fredkin gate with known pure-state inputs. 

\begin{equation}\label{eq:fredkintoU}
\begin{adjustbox}{height=10mm}
\begin{quantikz}[column sep=2mm]
\lstick{$\ket{\phi}$} &\ctrl{1} & \qw \\
\lstick{$\ket{\psi}$} &\swap{-1} & \qw \\
\lstick{$\ket{\pi}$} &\swap{-1}& \qw
\end{quantikz}
=\begin{quantikz}[column sep=2mm]
\lstick{$\ket{\phi}$} & \ctrl{1} & \ctrl{2}&\qw\\
\lstick{$\ket{\psi}$} & \gate{U} &\qw &\qw\\
\lstick{$\ket{\pi}$} &\qw & \gate{U^{-1}} &\qw
\end{quantikz}
\ifthen{\ket{\pi} = U\ket{\psi}}
\end{adjustbox}
\end{equation}

\subsection{Optimizing Qubit blocks}
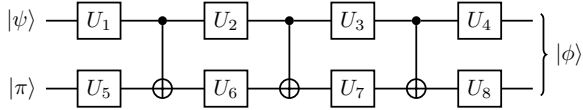
\begin{figure}[ht]
\centering
\begin{adjustbox}{width=0.9\linewidth}
\begin{quantikz}
\lstick{$\ket{\psi}$} &\gate{U_1} & \ctrl{1}&\gate{U_2} & \ctrl{1}&\gate{U_3} & \ctrl{1}& \gate{U_4}& \qw\rstick[wires=2]{$\ket{\phi}$} \\
\lstick{$\ket{\pi}$} &\gate{U_5} & \targ{}&\gate{U_6} & \targ{}&\gate{U_7} & \targ{}& \gate{U_8}& \qw
\end{quantikz}
\end{adjustbox}
\caption{Universal decomposition of two-qubit block $U$}
\label{fig:twoqubitblockdcompose}
\end{figure}

\begin{figure}[ht]
\centering
\vspace*{-4mm}
\begin{adjustbox}{width=0.6\linewidth}
\begin{quantikz}
\lstick{$\ket{\psi}$} &\gate{V_1} & \ctrl{1}&\gate{V_2}& \qw\rstick[wires=2]{$\ket{\phi}$}\\
\lstick{$\ket{\pi}$} &\gate{V_3} & \targ{}&\gate{V_4} & \qw
\end{quantikz}
\end{adjustbox}
\caption{Universal circuit for preparing quantum state $\ket{\phi}$}
\label{fig:twoqubitstateprepare}
\end{figure}
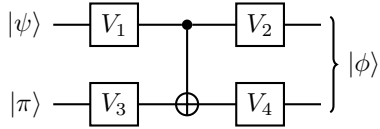

Our proposed QPO can be further generalized to optimize two-qubit blocks. A sequence of uninterrupted two-qubit gates is considered as a two-qubit block~\cite{qiskit}. The Qiskit transpiler optimizes these blocks by calculating the unitary matrices of these blocks and resynthesizing them using the KAK decomposition~\cite{kraus2001kak1, khaneja2001kak2}. Generally speaking, a two-qubit block can decomposed to a circuit consisting of at most three $\mathit{CNOT}$s and eight single-qubit gates~\cite{vidal2004universalthreeCNOT} as shown in Figure~\ref{fig:twoqubitblockdcompose}. If the two input qubit states $\ket{\psi}$ and $\ket{\pi}$ are in known pure states, the compiler can calculate the output state $\ket{\phi}$ statically. Existing research~\cite{mottonen2006dgateecomposition} has proved that any two-qubit state can be prepared by a $\mathit{CNOT}$ gate and four single-qubit gates. It means that we can substitute the two-qubit block with the state preparation circuit shown in Figure~\ref{fig:twoqubitstateprepare}. As a result, we reduce the number of $\mathit{CNOT}$ gates by two and the number of single-qubit gates by four.

The optimization of the two-qubit block can be generalized to n-qubit blocks. If we know the input state and the output state of an n-qubit block, we can use the state preparation circuit to substitute the original circuit. It has been proved that preparing a quantum state can require less $\mathit{CNOT}$ gates than preserving the unitary matrix~\cite{plesch2011statepreparation}. 

\section{Quantum state Analysis}

\label{sec:quantumstateanalysis}
\subsection{Basis-State Analysis}
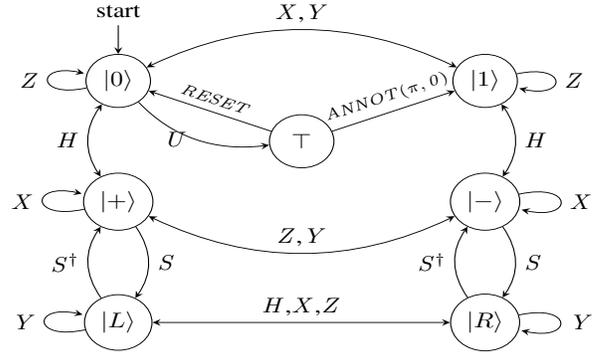
\begin{figure}[ht]
\centering
\begin{adjustbox}{width=0.9\linewidth,height = 4.7cm}
\begin{tikzpicture}[->, >=stealth]
\node[state, initial, initial where=above] at (-2.5, 0) (zero) {$\ket{0}$};
\node[state] at (2.5, 0) (one) {$\ket{1}$};
\node[state] at (2.5, -2) (minus) {$\ket{-}$};
\node[state] at (-2.5, -2) (plus) {$\ket{+}$};
\node[state] at (0, -1) (top) {$\top$};
\node[state] at (-2.5, -4) (left) {$\ket{L}$};
\node[state] at (2.5, -4) (right) {$\ket{R}$};
\draw
(one) edge[bend right, above, <->] node{$\mathit{X}$,$\mathit{Y}$} (zero)    
(minus) edge[bend left, above, <->] node{$\mathit{Z}$,$\mathit{Y}$} (plus)  
(one) edge[bend left, right, <->] node{$\mathit{H}$} (minus)    
(zero) edge[bend right, left, <->] node{$\mathit{H}$} (plus) 
(zero)  edge[bend right, left] node{$\mathit{U}$} (top)
(zero) edge[loop left] node{$\mathit{Z}$} (zero)
(one) edge[loop right] node{$\mathit{Z}$} (one)
(plus) edge[loop left] node{$\mathit{X}$} (plus)
(minus) edge[loop right] node{$\mathit{X}$} (minus)
(right) edge[loop right] node{$\mathit{Y}$} (right)
(left) edge[loop left] node{$\mathit{Y}$} (left)
(top) edge[right] node[sloped, above]{\scriptsize$\mathit{RESET}$} (zero)
(top) edge[right] node[sloped, above]{\scriptsize$\mathit{ANNOT(\pi,0)}$} (one)
(left) edge[above, <->] node{$\mathit{H}$,$\mathit{X}$,$\mathit{Z}$} (right)
(plus) edge[bend left, right, ->] node{$\mathit{S}$} (left)
(plus) edge[bend right, left, <-] node{$\mathit{S^\dagger}$} (left)
(minus) edge[bend left, right, ->] node{$\mathit{S}$} (right)
(minus) edge[bend right, left, <-] node{$\mathit{S^\dagger}$} (right)
;
\end{tikzpicture}
\end{adjustbox}
\caption{Partial automata for single-qubit basis-state analysis}
\label{fig:constant_transitions}
\end{figure}

For the purpose of applying QBO, we implemented a state automata, partially shown in Figure~\ref{fig:constant_transitions}, to track the basis state of each qubit. The automata consist in six distinguished single-qubit states: $\ket{0}$, $\ket{1}$, $\ket{+}$, $\ket{-}$, ${\ket{L}}$, and ${\ket{R}}$. 
Every single-qubit half- and quarter-turn gate transitions the states in the automata (not all of them reflected in Figure~\ref{fig:constant_transitions}).
Any other gate or operation (except special cases for $\mathit{SWAP}$, $\mathit{SWAPZ}$, and $\mathit{RESET}$), denoted as $\mathit{U}$, would change the state of the qubit to the \textit{unknown non-basis state} $\top$.
That is represented in the automata graph only with the ${\ket{0}}$ example but the same applies to all the other states.

As quantum processors are initialized in its lowest-energy state known as \textit{ground state}, all the qubits start in state $\ket{0}$.
The half- and quarter-turn gates transform basis states into basis states. For example, the Hadamard gate $\mathit{H}$ (a half-turn gate)  moves the Z-basis into the X-basis and vice versa.
The loop transitions in each node indicate a gate with no effect on that state, i.e. it is the eigenstate with an eigenvalue of 1, as explained in Section~\ref{sec:optSingleQubit}.

The instruction $\mathit{RESET}$ turns any state into the zero state.
In Figure~\ref{fig:constant_transitions}, that is illustrated as the only transition able to downgrade from $\top$ to $\ket{0}$. The annotation instruction $\mathit{ANNOT}$, to be discussed in subsection~\ref{subsec:annotation}, can transit between different states. Essentially, it can be used to change the state from $\top$ to a basis state, as exampled with the edge between $\top$ and $\ket{1}$.

The basis-state analysis also considers the effect of $\mathit{SWAP}$ and $\mathit{SWAPZ}$.
When they are encountered, the states of the involved qubits are swapped, including $\top$.


\subsection{Pure-State Analysis}

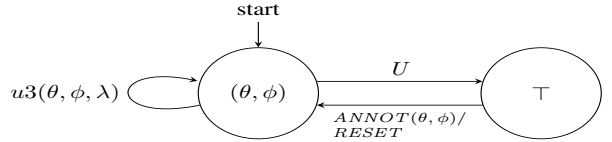
\begin{figure}[ht]
\centering
\vspace*{-6mm}
\begin{adjustbox}{width=0.9\linewidth,height = 2.0cm}
\begin{tikzpicture}[>=stealth]
\node[state, initial, initial where=above,minimum size=1.7cm] at (-2, 0) (pure) {$(\theta,\phi)$};
\node[state,minimum size=1.7cm] at (2, 0) (top) {$\top$};
 \draw
 (pure.15)  edge[above, ->] node{$\mathit{U}$} (top.165)
 (pure) edge[loop left] node{$u3(\theta,\phi,\lambda)$} (pure)
 (top.195) edge[below, ->] node{\scriptsize \begin{tabular}{l}
    $\mathit{ANNOT(\theta, \phi)/}$\\
    $\mathit{RESET}$
\end{tabular}} (pure.345);
\end{tikzpicture}
\end{adjustbox}
\caption{Automata for pure-state analysis}
\label{fig:pure_transitions}
\end{figure}

As QBO, QPO requires a pure-state analysis to track the single-qubit pure states. A single qubit is in pure state when it is not entangled with the other qubits. 
%
Any single-qubit pure state can be represented by a state vector with two parameters $\theta$ and $\phi$, $\ket{\psi(\theta,\phi)} = cos(\frac{\theta}{2})\ket{0} + e^{i\phi}sin(\frac{\theta}{2})\ket{1}$. Knowing the pure state $\ket{\psi(\theta,\phi)}$, we can use a single-qubit gate $u3(\theta, \phi, 0)$ to generate this state from $\ket{0}$ state, $\ket{\psi(\theta,\phi)} = u3(\theta, \phi, 0)\ket{0}$. Therefore, we choose these two parameters to record the single-qubit pure state information. In our analysis, each qubit associates a tuple ($\theta$, $\phi$). 
When the qubit is not in pure state, the parameters are set to $\top$. 

The tuple $(\theta,\phi)$ is updated to track the pure state information for each qubit.
Since any single-qubit gate can be expressed by the $u3(\theta,\phi,\lambda)$ gate, when we apply a $u3(\theta,\phi,\lambda)$ gate to a qubit in the pure state $(\theta_0,\phi_0)$, the output state would be a pure state $(\theta_1,\phi_1)$. Calculating the parameters $(\theta_1,\phi_1)$ of the output state is analogous to merging two $u3$ gates, since $\ket{\psi(\theta_1,\phi_1)} = u3(\theta,\phi,\lambda)\ket{\psi(\theta_0,\phi_0)}$ = $u3(\theta,\phi,\lambda)$ $u3(\theta_0,\phi_0, 0)\ket{0}$ = $u3(\theta',\phi',\lambda')\ket{0}$ = $u3(\theta_1,\phi_1,0)\ket{0}$. Since the $\lambda'$ parameter does not change the $\ket{0}$ state and can be ignored, we have $\theta_1 = \theta’$ and $\phi_1 = \phi’$. In our implementation, we leverage the gate merging function in Qiskit to calculate the output state parameters $\theta_1$ and $\phi_1$.

When a multi-qubit gate is applied to the qubits in pure state, the output might be in mixed state. Therefore, the resulting states are marked as $\top$ for each implicated qubit. The $\mathit{RESET}$ instruction will reset the qubit back to ground state $(0, 0)$. The $\mathit{ANNOT(\theta, \phi)}$ annotation, to be discussed in subsection~\ref{subsec:annotation}, will transform the qubit to pure state $(\theta, \phi)$. The transitions among these states are illustrated by the automata in Figure~\ref{fig:pure_transitions}. Similar to the basis-state analysis, our pure state analysis considers $\mathit{SWAP}$ and $\mathit{SWAPZ}$ gates.
When they are encountered, the pure states of the involved qubits are swapped, including $\top$.

Section~\ref{sec:fromConsttoPure} discusses the optimization for a $\mathit{SWAP}$ gate with two known pure states. The optimization needs the unitary gates that transform one pure state to the other. With the two-parameter ($\theta$, $\phi$) representation, it is easy to generate such unitary gates. The gate $u3(\theta_2-\theta_1, \phi_2-\phi_1)$ transforms the pure state $\ket{\psi(\theta_1, \phi_1)}$ to $\ket{\psi(\theta_2, \phi_2)}$, $\ket{\psi(\theta_2, \phi_2)} = u3(\theta_2-\theta_1, \phi_2-\phi_1)\ket{\psi(\theta_1, \phi_1)}$. 

\subsection{State Annotation}
\label{subsec:annotation}
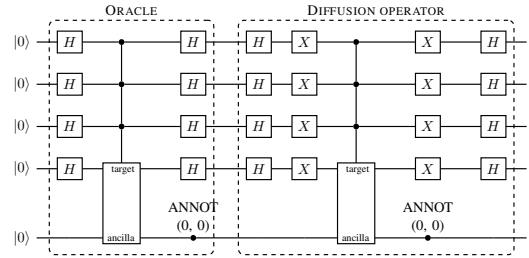
\begin{figure}
\centering
\begin{adjustbox}{width=0.8\linewidth}{
\begin{quantikz}
\lstick{$\ket{0}$} &\gate{H}\gategroup[5,steps=3,style={dashed,rounded corners, inner xsep=2pt},background]{{\sc Oracle}} & \ctrl{1} & \gate{H} & \gate{H}\gategroup[5,steps=5,style={dashed,rounded corners, inner xsep=2pt},background]{{\sc Diffusion operator}} & \gate{X} & \ctrl{1} & \gate{X} & \gate{H} &\qw \\
\lstick{$\ket{0}$} &\gate{H} & \ctrl{1} & \gate{H} & \gate{H} & \gate{X} & \ctrl{1} & \gate{X} & \gate{H} &\qw \\
\lstick{$\ket{0}$} &\gate{H} & \ctrl{1} & \gate{H} & \gate{H} & \gate{X} & \ctrl{1} & \gate{X} & \gate{H} &\qw \\
\lstick{$\ket{0}$} &\gate{H}& \gate[wires=2][0.9cm]{} \gateoutput{target} &\gate{H} & \gate{H} & \gate{X} & \gate[wires=2][0.9cm]{} \gateoutput{target} & \gate{X} & \gate{H} &\qw \\
\lstick{$\ket{0}$} &\qw & \gateoutput{ancilla} & \phase[label position=above]{\begin{array}{c} \text{ANNOT} \\ \text{(0, 0)} \\ \text{} \end{array}} &\qw & \qw & \gateoutput{ancilla} & \phase[label position=above]{\begin{array}{c} \text{ANNOT} \\ \text{(0, 0)} \\ \text{} \end{array}} & \qw &\qw 
\end{quantikz}
}
\end{adjustbox}
\caption{4-qubit Grover's algorithm using multi-controlled Z gates with ``clean'' ancilla qubits and annotations}
\label{fig:grover}
\end{figure}

Determining whether a generic quantum state is entangled is an NP-hard problem~\cite{gurvits2003entanglementNPHard}. In general, it is hard to infer information about the states from a quantum circuit using a classical machine efficiently. However, based on the understanding of the quantum program, the programmer can provide information to facilitate state analysis. For example, in quantum networks for elementary arithmetic operations~\cite{vedral1996arithmatic}, the network uses reverse computation to unentangle and reuse qubits. The programmers know these qubits are unentangled after reverse computation and they can annotate that these qubits are in particular pure states. Another example is, ``clean" ancilla qubits are commonly used in quantum computing. The ``clean" ancilla qubits are in $\ket{0}$ state and can be reused after the gate. As shown in Figure~\ref{fig:grover}, after the multi-controlled Toffoli gate, the ancilla qubit remains to be in state $\ket{0}$. To leverage such user-level information, we introduce annotations, which inform the compiler that the qubits are in certain quantum states. For example, in the pure state analysis, we use the annotation $\mathit{ANNOT}(\theta, \phi)$ to indicate a quantum state is in pure state $\ket{\psi(\theta, \phi)}$. The programmer can insert annotations based on the understanding of the quantum program. The annotations can also be inserted by the compiler automatically. For example, when the programmer uses the gate design with ``clean'' ancilla qubits, the compiler may automatically insert annotations $\mathit{ANNOT}(0, 0)$ for such ancilla qubits.

By introducing the annotations, we can avoid the complex quantum state analysis and improve the scalability of our proposed optimization pass. 

\section{Methodology}
\label{sec:methodology}
\subsection{Compiler Implementation}
We implemented our QBO and QPO passes on the open-source quantum computing framework Qiskit 0.18~\cite{qiskit} and our implementation is publicly available~\footnote{\texttt{\url{https://github.com/1ucian0/rpo}}~\cite{rpocode}}.
Qiskit organizes the transpilation passes in pass managers and it currently includes four pass managers for each level of optimization.
The level 3 provides the maximal optimization at the expense of longer transpilation time.
As part of our implementation, we extended the pass manager for level 3 to include our optimizations.

Figure~\ref{fig:passmanager} outlines the sequence of passes in level 3 and the additions (underlined) that we introduced.
The input circuit is run through the \texttt{QBO} pass first.
The effect of this early optimization cascades in the rest of the pass manager, since any reduction in the gate count will improve the speed and effectiveness of subsequent passes. 
Additionally, \texttt{QBO} checks basis states of the $\mathit{SWAPZ}$ gates in the input circuit, if there are any.
If the condition in Equation~\ref{eq:swapzZero} does not hold for a specific $\mathit{SWAPZ}$ gate, the gate is decomposed into 2 $\mathit{CNOT}$ gates, following the definition from Equation \ref{eq:swapz}. This guarantees that the $\mathit{SWAPZ}$ gates from this point on are semantically equivalent to $\mathit{SWAP}$ gates.
After the $\mathit{SWAP}$ gates are inserted during the routing process (line 4) a new pass of \texttt{QBO} (line 5) optimizes those inserted $\mathit{SWAP}$ gates.
In line 6 and 7, we reuse existing Qiskit functionalities, \texttt{Unroller} and \texttt{Optimize1qGate} to prepare for the \texttt{QPO} pass (line 8).
The \texttt{Unroller} pass decomposes all the circuit gates into the list of gates defined by the parameter.
The variable \texttt{basis\_gates} is a list of the primitive gates supported by the quantum device. In line 8, the list is extended with the gates $\mathit{SWAP}$ and $\mathit{SWAPZ}$, since \texttt{QPO} understands them.
The \texttt{Optimize1qGates} pass merges the single-qubit gates into a single unitary gate. 
After \texttt{QPO}, the circuit is optimized in a loop until a fixed point is reached. 
This loop is expensive and we decided to place \texttt{QBO} and \texttt{QPO} out of it.
The loop iterates at least twice in order to find the fixed point. The optimizations in the loop (line 10) do not modify the state invariant on the qubits. Therefore, there is no gain running \texttt{QBO/QPO} more than once.

\begin{figure}
\centering
\begin{adjustbox}{height=17mm}
\begin{lstlisting}
<@\underline{QBO()}@>
Unroller(basis_gates)
<layout selection>
<routing process>
<@\underline{QBO()}@>
<@\underline{Unroller(basis\_gates + swap + swapz)}@>
<@\underline{Optimize1qGates()}@>
<@\underline{QPO()}@>
while not <fixed point>{
  <optimizations>}
\end{lstlisting}
\end{adjustbox}
\caption{Optimization level 3 in Qiskit 0.18 (RPO additions underlined)}
\vspace*{-4mm}
\label{fig:passmanager}
\end{figure}

\subsection{Benchmarks and System configuration}

To evaluate our proposed compiler optimization, we run our experiments upon the following algorithms:

\textbf{Bernstein-Vazirani algorithm}: A blackbox function $f(x)$ is guaranteed to be the dot product between $x$ and a bit string $s$: $f(x) = x\cdot s$. Given an oracle that implements $f(x)$, the algorithm finds the hidden bit string $s$ with a single evaluation.


\textbf{Quantum Phase Estimation (QPE) algorithm}: QPE estimates the phase of an eigenvector of a unitary matrix. Given a quantum state $\ket{\psi}$ which is the eigenvector of a unitary matrix $U$, $U\ket{\psi} = e^{2\pi i\theta}\ket{\psi}$, QPE estimates the phase $\theta$.

\textbf{VQE algorithm}: Variational Quantum Eigensolver (VQE) is a hybrid quantum/classical algorithm which finds the eigenvalues of a matrix $\mathit{H}$. 
In the VQE algorithm, the circuit for preparing the quantum state is called \emph{anstaz}.
We use the VQE program and the hardware-efficient ansatz $\mathit{RY}$ from Qiskit Aqua~\cite{qiskit}. In our experiment, we use the VQE algorithm to solve the Max-Cut problem~\cite{garey1979maxcut}.

\textbf{Quantum Volume}: Quantum volume~\cite{cross2019quantumvolume} is a metric for characterizing quantum system performance. It is calculated by taking various quantum computer features into account, such as gate error, and connectivity. The quantum volume circuit is randomly generated with a fixed but generic form.

\textbf{Grover's search algorithm}: Given a set $X$ of $N$ elements and a boolean function $f: X\rightarrow {0,1}$, Grover's algorithm finds an element $x_i$ in $X$ such that $f(x_i) = 1$.

\begin{figure}[h]
    \centering
    \begin{adjustbox}{height=18mm}
    \begin{subfigure}{0.5\linewidth}
    \centering
        \includegraphics[width=0.9\linewidth]{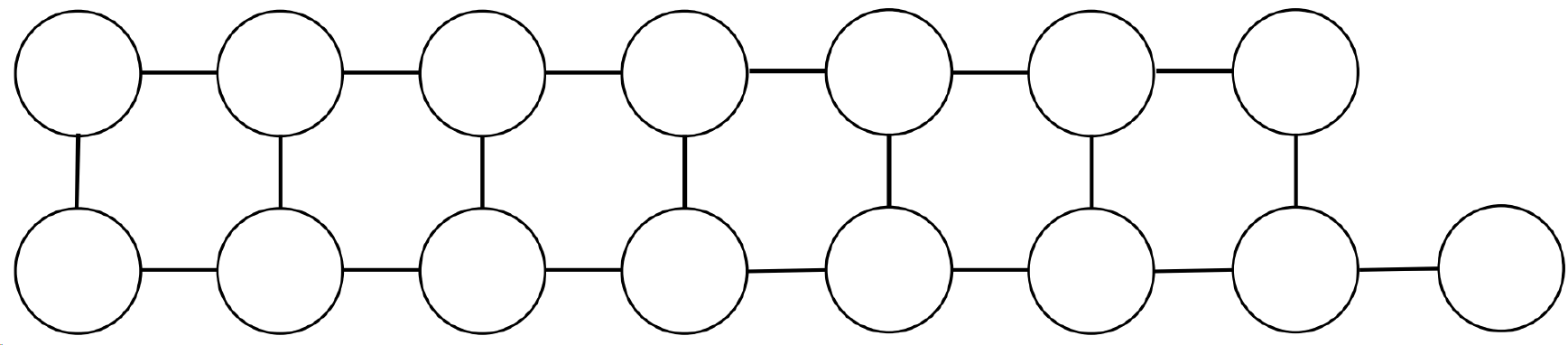}
        \caption{\texttt{ibmq\_16\_melbourne}}
        \includegraphics[width=0.8\linewidth]{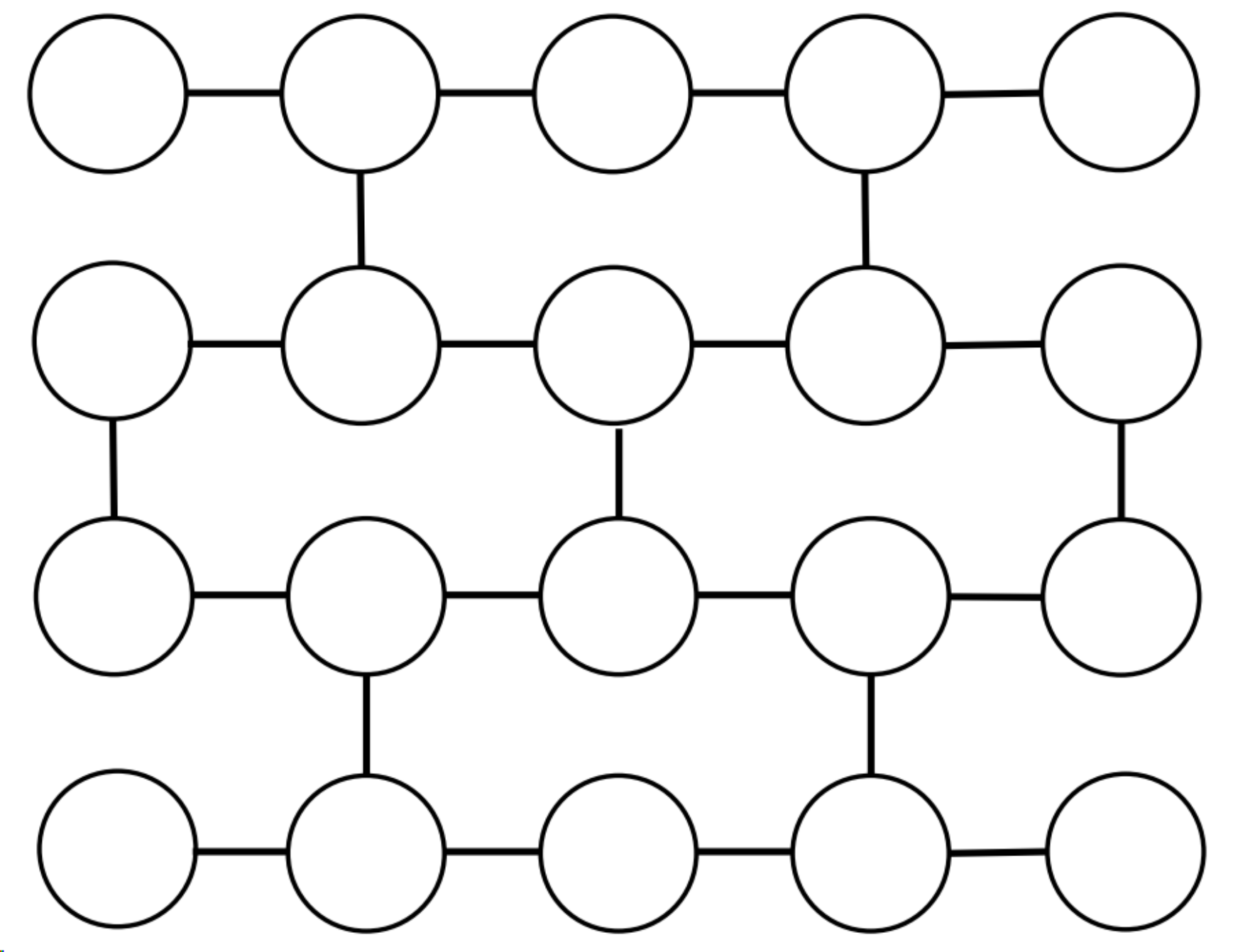}
        \caption{\texttt{ibmq\_almaden}}
    \end{subfigure}%
    \begin{subfigure}{0.5\linewidth}
    \centering
        \includegraphics[width=0.9\linewidth]{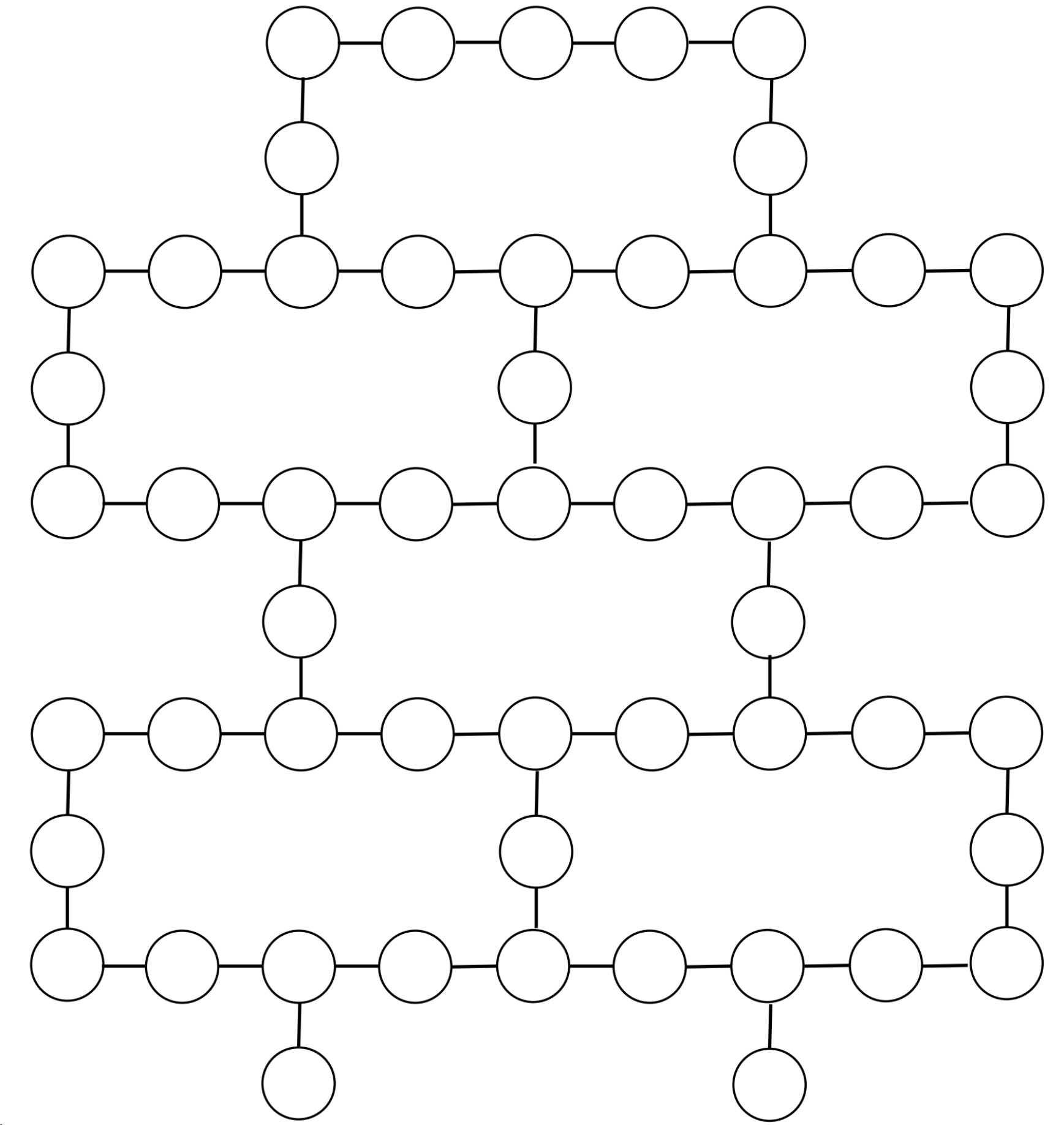}
        \caption{\texttt{ibmq\_rochester}}
        \label{subfig:rochester}
    \end{subfigure}
    \end{adjustbox}
    \caption{Connectivity map of three different IBM quantum computers}
        \vspace*{-4mm}
    \label{fig:connectivity}
    \end{figure}

Since the $\mathit{RESET}$ is currently not supported by the IBMQ quantum hardware, none of the circuits used in our experiments include the $\mathit{RESET}$ instruction. We run our experiments with connectivity maps and noise properties from three different quantum computers, a 15-qubit quantum computer \texttt{ibmq\_16\_melbourne}, a 20-qubit quantum computer \texttt{ibmq\_almaden}, and a 53-qubit quantum computer \texttt{ibmq\_rochester}.
The connectivity maps of these three quantum computers are shown in Figure~\ref{fig:connectivity}. We compare our optimization pass with the hoare logic pass~\cite{qiskit_hoare_logic} implemented in the Qiskit transpiler. For a fair comparison, we append the hoare logic pass to the level 3 pass manager. We also run our experiments on these real quantum computers to evaluate the fidelity rate improvement from our compiler optimization. 

Each circuit was transpiled several times to mitigate the effect of corner cases given the non-deterministic nature of Qiskit transpiler.
For example, the \texttt{StochasticSwap} routing pass included in Qiskit returns significantly different results depending on the random seed and the input circuit.
The reported $\mathit{CNOT}$ gate count and the transpilation time are the medians of twenty-five (25) transpilation results. The reported median $\mathit{CNOT}$ gate count is very close to the average $\mathit{CNOT}$ gate count. We use the geometric mean to calculate the average ratio of $\mathit{CNOT}$ gate reduction.

\section{Performance}
\label{sec:performance}

In this section, we first show a case study on the Bernstein-Vazirani algorithm, for which our optimization pass will optimize the boolean oracle with $\mathit{CNOT}$ gates into phase oracle with single-qubit gates.  Then, we provide case studies on four widely used algorithms namely the quantum phase estimation (QPE) algorithm, variational quantum eigensolver (VQE) algorithm, quantum volume benchmark, and Grover's algorithm. Subsequently, we show the annotations improve the scalability of our optimization. Next, we study the impact of the backend connectivity on the optimization. In the end, we run the experiments on real quantum computers to show the success rate improvement with our optimization. 

\begin{table*}[htbp]
  \centering
  \small
    \resizebox{\linewidth}{!}{%
  \begin{tabular}{|c|c|c|c|c|c|c|c|c|c|c|c|c|c|c|c|c|c|c|c|c|c|c|c|c|}
    \hline
    & \multicolumn{6}{c|}{QPE}
   & \multicolumn{6}{c|}{VQE}
   & \multicolumn{6}{c|}{Quantum Volume} &\multicolumn{6}{c|}{Grover} \\
     \hline
    Metric & \multicolumn{3}{c|}{$\mathit{CNOT}$ gate count} &  \multicolumn{3}{c|}{transpile time(s)}& \multicolumn{3}{c|}{$\mathit{CNOT}$ gate count} & \multicolumn{3}{c|}{transpile time(s)}& \multicolumn{3}{c|}{$\mathit{CNOT}$ gate count} & \multicolumn{3}{c|}{transpile time(s)}
    & \multicolumn{3}{c|}{$\mathit{CNOT}$ gate count} & \multicolumn{3}{c|}{transpile time(s)}\\
    \hline
    Optimization & level3 &hoare& RPO& level3 &hoare& RPO & level3 &hoare& RPO & level3 &hoare& RPO & level3 &hoare& RPO & level3 &hoare& RPO& level3 &hoare& RPO & level3 &hoare& RPO\\
    \hline
    4-qubits & 24 & 21 & 18 & 0.29 & 0.41 & 0.28 & 56 & 51 & 47 & 0.43 & 0.77 & 0.42 & 38 & 28 & 26 & 0.39 & 0.74 & 0.39& 168 & 159 & 157 & 1.80 & 2.14 & 1.51 \\
    \hline
    6-qubits & 66 & 62 & 54 & 0.81 & 1.00 & 0.73 & 147 & 141 & 136 & 1.53 & 1.62 & 1.38 & 75 & 75 & 72 & 1.73 & 1.93 & 1.51& 359 & 345 & 322 & 4.61 & 5.72 & 4.78 \\
    \hline
    8-qubits& 124 & 117 & 106 & 1.34 & 1.74 & 1.24 & 301 & 289 & 285 & 1.95 & 2.71 & 1.99 & 165 & 158 & 147 & 2.92 & 3.85 & 3.27 & 1551 & 1491 & 1463 & 16.6 & 30.2 & 13.8 \\
    \hline
    10-qubits & 205 & 197 & 172 & 1.88 & 2.30 & 1.59 & 485 &470 & 459 & 2.77 & 4.78 & 2.61 & 327 & 313 & 282 & 6.39 & 7.15 & 5.04  & 6358 & 6309 & 6275 & 52.4 & 303.7 & 45.9 \\
    \hline
    12-qubits & 268 & 261 & 225 & 2.77 & 4.39 & 3.11 & 720 & 699 & 683 & 4.74 & 6.76 & 4.37 & 429 & 424 & 399 & 7.56 & 11.84 & 7.36 & 25386 & 25254 & 25008 & 232.5 & 10271.8 & 231.8 \\
    \hline
    14-qubits & 500 & 500 & 451 & 4.95 & 10.98 & 7.04 & 1142 & 1136 & 1136 & 5.74 & 11.72 & 5.69 & 1505 & 1491 & 1479 & 19.62 & 25.94 & 16.27 & 101020 & N.A. & 100762 & 1769.4 & N.A. & 1828.2\\
    \hline
    
  \end{tabular}
  }
  \caption{Median of $\mathit{CNOT}$ gates and transpilation time of three quantum algorithms with different size (on \texttt{ibmq\_16\_melbourne})}
    \vspace*{-4mm}
  \label{table:allbench}
\end{table*}

\subsection{Bernstein-Vazirani Algorithm}
Circuits implementing the Bernstein-Vazirani Algorithm are used extensively for benchmarking in recent quantum computing researches~\cite{tannu2019ensemble, tannu2019notallqubits, tannu2019mitigatingbias}.
QBO has a particular effect on this algorithm implementations that is noteworthy.

\begin{figure}[h]
        \vspace*{-4mm}
    \captionsetup[subfigure]{position=b}
    \centering
    \begin{subfigure}[t]{0.5\linewidth}
        \vskip 0pt 
    \centering
        \includegraphics[width=0.8\linewidth, height=35mm]{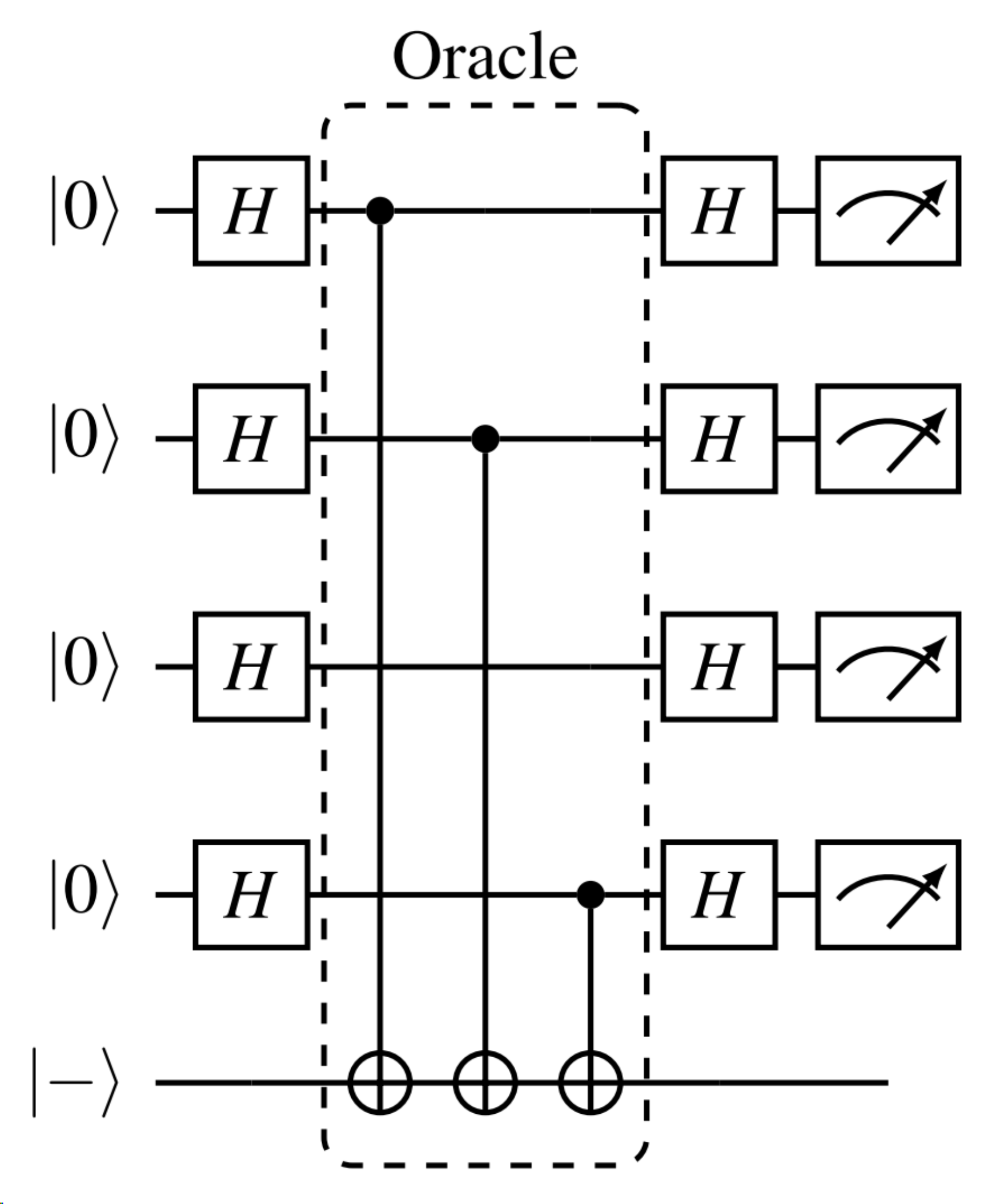}
        \vspace*{-2mm}
        \caption{Boolean oracle}
        \label{subfig:BVancilla}
    \end{subfigure}%
    \begin{subfigure}[t]{0.5\linewidth}
    \vskip 0pt 
    \centering
        \includegraphics[height=30mm]{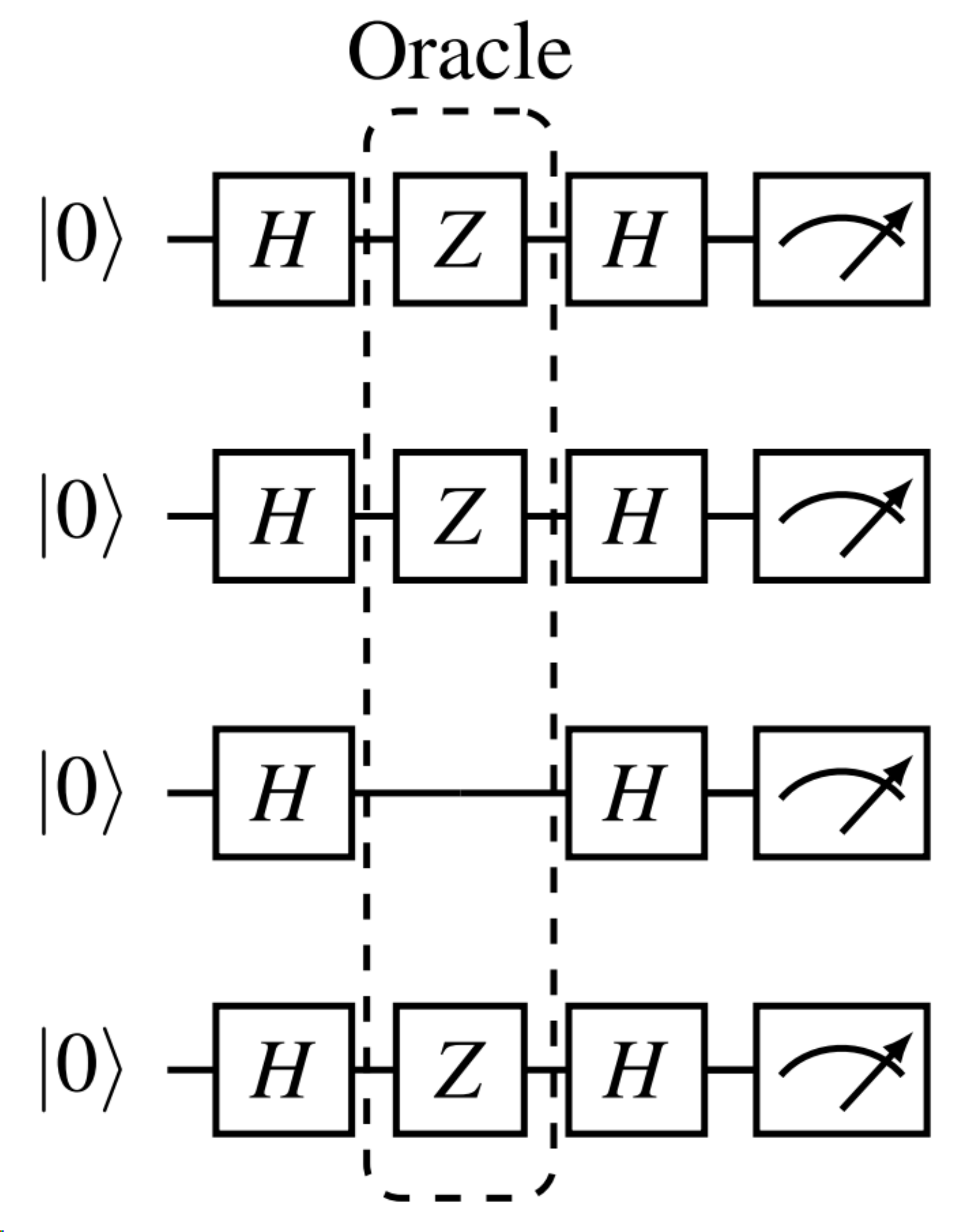}
        \vspace{10pt} 
        \caption{Phase oracle}
        \label{subfig:BVphaseoracle}
    \end{subfigure}
    \vspace*{-2mm}
    \caption{Two different circuits for Bernstein-Vazirani algorithm}
    \vspace*{-4mm}
    \label{fig:BValgorithm}
    \end{figure}

The oracle that implements the function $f(x)$ can be represented in two different ways.
The boolean oracle method converts an irreversible computation to a reversible one~\cite{bennett1973logicalreversibility}. The other method is to use phase oracles~\cite{kenigsberg2006phaseoracle}, encoding  $f(x)$ into phase amplitudes. The phase oracle for the Bernstein-Vazirani algorithm can be done with only $\mathit{Z}$ gates~\cite{du2001BVphase}. Figure~\ref{fig:BValgorithm} shows two different implementations of 4-qubit Bernstein-Vazirani algorithm with hidden bit string $s = 1011$.

The first design requires an extra ancilla qubit and $\mathit{CNOT}$ gates. The second design only includes single-qubit gates and it is more feasible for noisy quantum systems.
Notice that the ancilla qubit is in the $\ket{-}$ state. Following the discussion in Section~\ref{subsec:constantopt}, our QBO pass substitutes the $\mathit{CNOT}$ gates with $\mathit{Z}$ gates and the optimized circuit is the same as the design with phase oracle. In other words, QBO converts the design in Figure~\ref{subfig:BVancilla} into Figure~\ref{subfig:BVphaseoracle}. 

We found that our optimization can optimize the costly boolean oracle to a design which has the same cost as the phase oracle. Besides the Bernstein-Vazirani algorithm, the boolean oracles for the Grover's algorithm~\cite{figgatt2017groverboolean} and general cases~\cite{nielsen2002quantumcomputation} can also be optimized by our pass.
In comparison, such boolean oracles can't be optimized by the Qiskit compiler or the hoare logic pass.  

\subsection{Quantum Algorithms}

In this section, we consider four practical quantum algorithms: quantum phase estimation, VQE, quantum volume, and Grover's search algorithm.
We compare RPO against the Qiskit compiler with optimization level 3 and the optimization level 3 with hoare logic pass, using the backend properties from \texttt{ibmq\_16\_melbourne}, which has 15 qubits. The median of $\mathit{CNOT}$ gate count and transpilation time are shown in Table~\ref{table:allbench}. For all of the circuits, the resulting $\mathit{CNOT}$ gate count of our pass manager is less than or equal to that of level 3. For most of the circuits, the compilation time of our pass manager is shorter than that of level 3, even though we included extra optimization passes. This is due to the early QBO, which cascades its effect to the rest of the passes in the pass manager. Since any reduction in the gate count will improve the subsequent passes, the overall compile time can be reduced. Our RPO pass results in more efficient circuit design and less compile time compared to the hoare logic pass. By checking the optimized circuit, we found that all the gates that are optimized by the hoare logic pass can be captured by our RPO pass. The median of single-qubit gate count and circuit depth are shown in Table~\ref{table:depthsinglegate} in Appendix~\ref{appendix:additional results}. As we can see from Table~\ref{table:depthsinglegate}, both the single-qubit gate count and circuit depth are improved as a result of our optimization.

For the QPE algorithm, when the logical circuit is decomposed to the basic gates, some of the $\mathit{CNOT}$ gate can be optimized by our compiler pass. Therefore, our optimized circuits have lower $\mathit{CNOT}$ gate count for all different numbers of qubits. Notice that our optimization has a significant impact for the shallow circuits. For the 4-qubit QPE algorithm, our optimization reduced the $\mathit{CNOT}$ gate count by $25\%$. On average, our optimization leads to $18.0\%$ decrease in the $\mathit{CNOT}$ gate count and $5.5\%$ decrease in the transpilation time for the QPE algorithm.

For the VQE algorithm, we use the hardware-efficient ansatz RY as the circuit design. The hardware-efficient ansatz is concise which limits the possible optimizations. However, when mapped to the physical qubits, the compiler introduces extra $\mathit{SWAP}$ gates. Therefore, it is still possible to optimize the circuit. As the number of qubit increases, the number of $\mathit{CNOT}$ gate optimized by our pass also increases. However, when the qubit count is close to the total number of qubits in the device (for this case 15), all the qubit will quickly fall into the non-basis/pure state $\top$, and our optimization only optimized a small amount of $\mathit{CNOT}$ gates. In the best case, our optimization reduced the $\mathit{CNOT}$ gate count by $16\%$ for the 4-qubit VQE algorithm. Our optimization leads to an average of $5.8\%$ decrease in the $\mathit{CNOT}$ gate count and $7.7\%$ decrease in the transpilation time for VQE algorithm.

For the quantum volume benchmark, since it is a randomly generated benchmark, the qubits are entangled and it is difficult to analyze the quantum states. Nevertheless, our optimization remains effective. 

Since the long compile time may cause compilation failure, we only compile one iteration of the Grover's algorithm. In the 14-qubit case, the hoare logic pass failed due to long compilation time. On average, our optimization reduces the $\mathit{SWAP}$ gate count by $2.4\%$ and the transpilation time by $7.3\%$.


Across these four benchmarks, our optimization reduces the $\mathit{CNOT}$ gate count by $11.7\%/4.5\%$ and the transpilation time by $7.1\%/40.0\%$ on average compared to Qiskit level 3 and Hoare logic, respectively.

\begin{figure*}
\centering 
    \begin{subfigure}{0.3\textwidth}
        \includegraphics[width=\columnwidth, height=35mm]{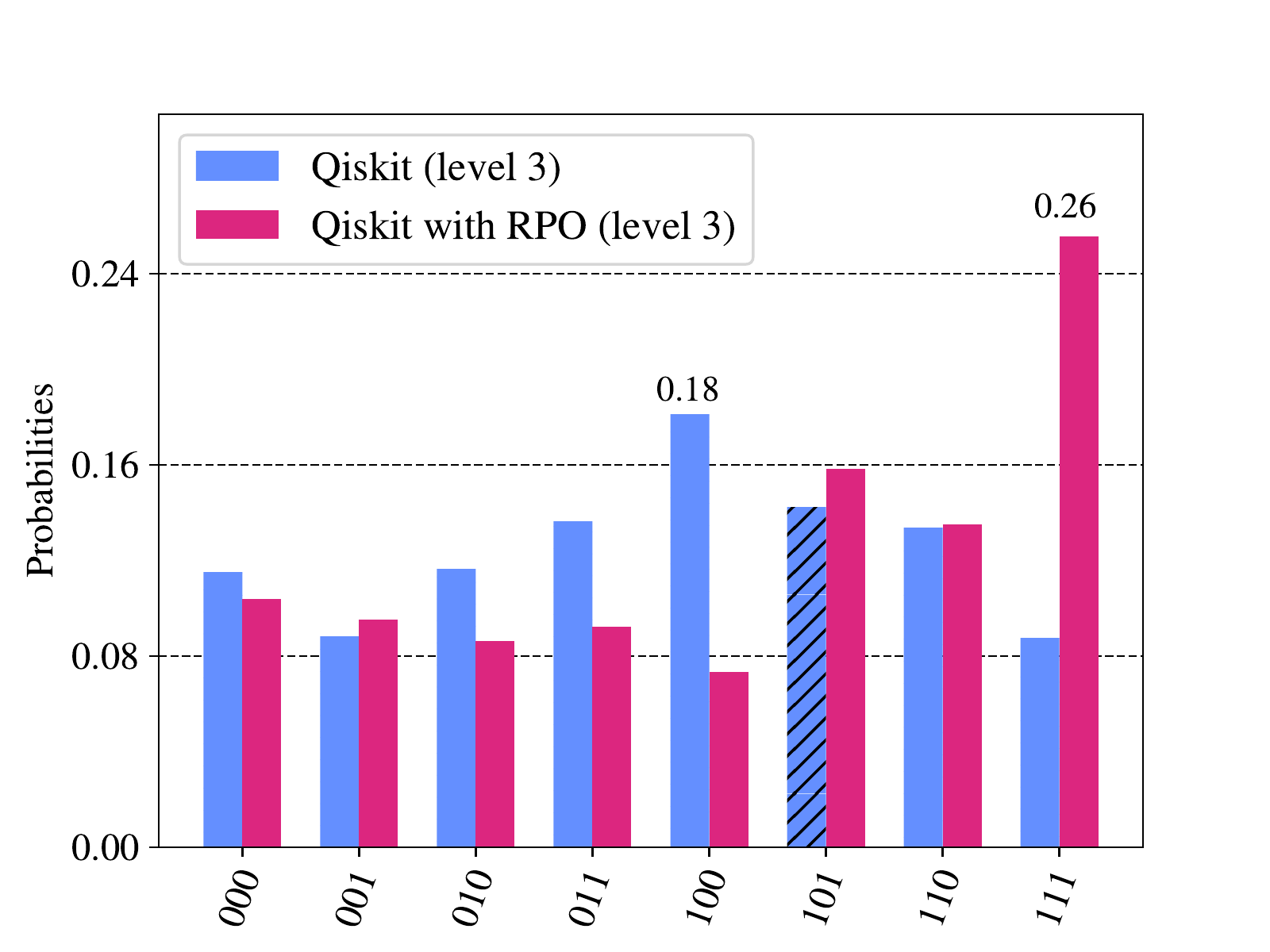}
        \caption{\texttt{ibmq\_16\_melbourne}}
    \end{subfigure}
    \begin{subfigure}{0.3\textwidth}
        \includegraphics[width=\columnwidth, height=35mm]{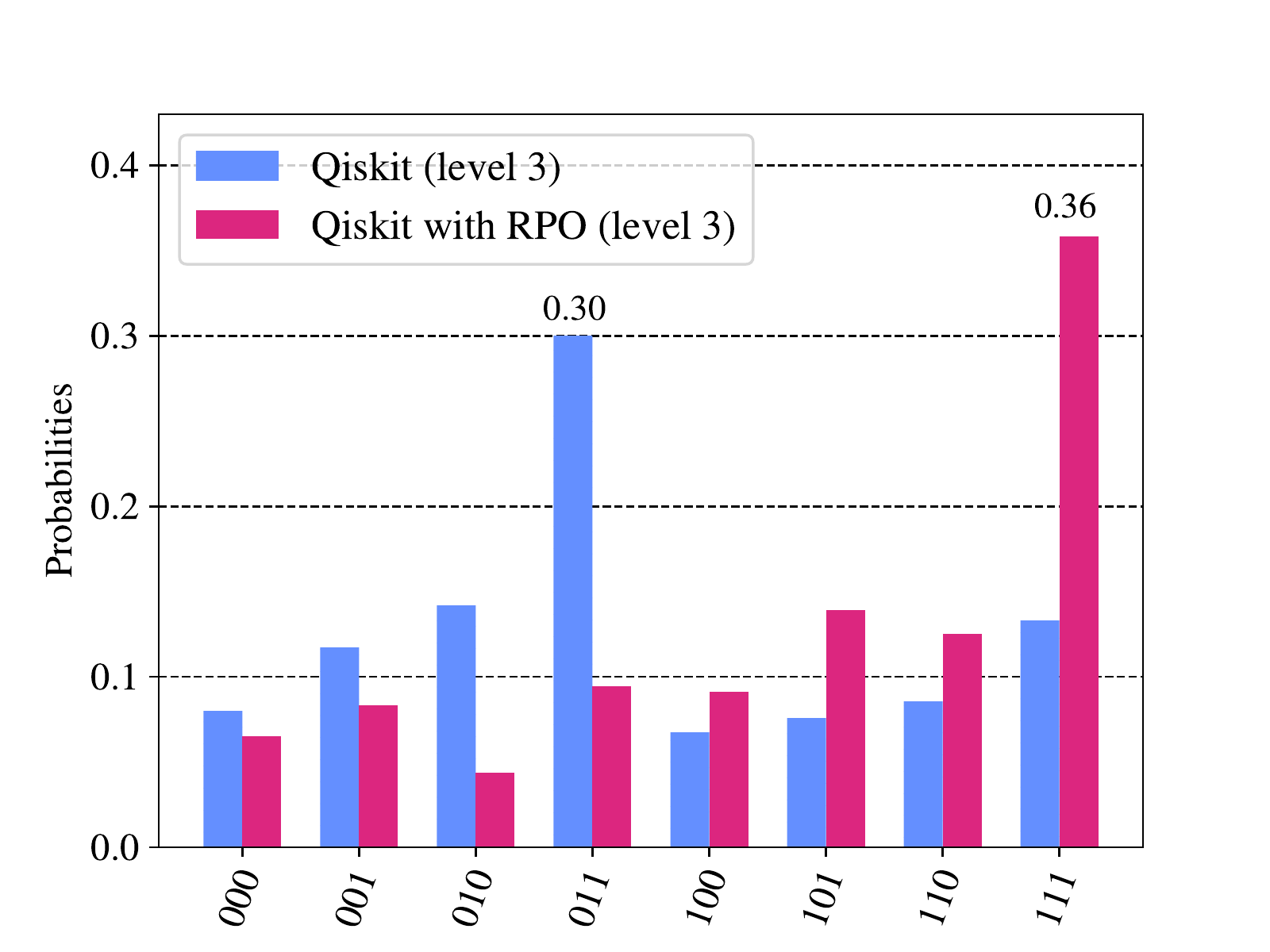}
        \caption{\texttt{ibmq\_almaden}}
    \end{subfigure}
        \begin{subfigure}{0.3\textwidth}
        \includegraphics[width=\columnwidth, height=35mm]{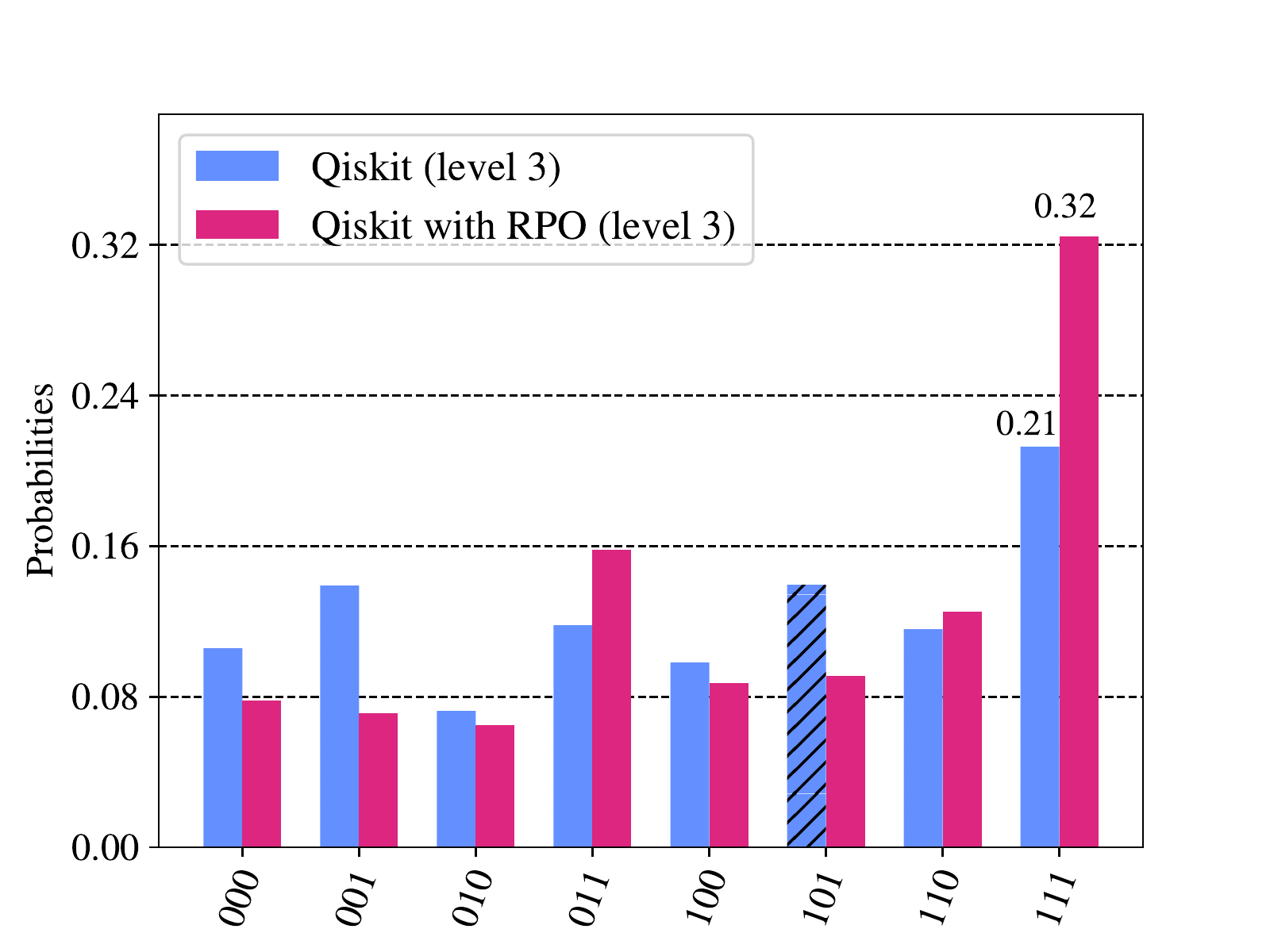}
        \caption{\texttt{ibmq\_rochester}}
    \end{subfigure}
\caption{Output distribution of QPE algorithm on three different quantum computers}
\vspace*{-6mm}
\label{fig:QPE_realdevice}
\end{figure*}

\subsection{Quantum Algorithm with Annotations}
\label{sec:annotation}
In this section, we use the Grover's algorithm to demonstrate that annotations can significantly improve the scalability of our optimization. It is a common practice to use quantum gates with ancilla qubits. Introducing ancilla qubits can significantly reduce the circuit size. For example, when using the multi controlled Toffoli gate without ancilla qubits, the 8-qubit grover's algorithm circuit consists of approximately $\sim1500$ $\mathit{CNOT}$ gates. We can use another design which requires six ``clean'' ancilla qubits. The ancilla qubit design only consists of approximately $\sim400$  $\mathit{CNOT}$ gates. Similar to Figure~\ref{fig:grover}, we can add annotations for the ``clean'' ancilla qubit. 

The n-qubit Grover's algorithm requires $O(\sqrt{2^n})$ iterations to maximize the probability amplitude of the correct output. Each iteration consists of an oracle and a diffusion operator. We use the multi-controlled Toffoli gate with ancilla qubits and test the 8-qubit Grover's algorithm with different number of iterations. The experiment results of different optimization passes are shown in Table~\ref{table:groverlength}. Without annotations, after first few iterations, all the qubits will fall into the non-basis/pure state $\top$. Therefore, $\mathit{CNOT}$ gate count reduced by RPO is $\sim$ $30$ regardless of the number of iterations. By introducing annotations, the qubits can transfer from $\top$ back to basis/pure state. 
For the 2-iteration case, our optimization reduced the $\mathit{CNOT}$ gate count by $10.8\%$. When the number of iteration is greater than eight, our optimization reduced a constant fraction $\sim$ $7.4\%$ of the $\mathit{CNOT}$ gate count. 

\begin{table}[bhtp]
  \centering
  \small
    \resizebox{\columnwidth}{!}{%
  \begin{tabular}{|c|c|c|c|c|c|c|c|c|c|c|c|}
     \hline
    Metric & \multicolumn{3}{c|}{$\mathit{CNOT}$ gate count} & \multicolumn{3}{c|}{depth} & \multicolumn{3}{c|}{transpile time(s)}\\
    \hline
    Optimization & level3 & RPO & RPO w/ Annot & level3 & RPO & RPO w/ Annot & level3 & RPO & RPO w/ Annot\\
    \hline
    2-iteration & 653 & 625 & 583 & 626 & 619 & 600 & 7.37 & 7.15 & 6.84\\
    \hline
    4-iteration & 1315 & 1280 & 1187 & 1249 & 1238 & 1200 & 15.11 & 14.88 & 13.89\\
    \hline
    6-iteration & 1882 & 1847 & 1709 & 1826 & 1815 & 1748 & 21.41 & 21.78 & 19.59\\
    \hline
    8-iteration & 2559 & 2527 & 2362 & 2443 & 2435 & 2350 & 25.74 & 29.20 & 27.45\\
    \hline
    10-iteration & 3111 & 3079 & 2886 & 3005 & 2994 & 2897 & 37.49 & 35.65 & 32.61\\
    \hline
    12-iteration & 3695 & 3660 & 3419 & 3606 & 3595 & 3478 & 45.51 & 45.02 & 43.09\\
    \hline
    14-iteration & 4288 & 4251 & 3979 & 4192 & 4179 & 4051 & 48.67 & 48.12 & 48.02\\
    \hline
  \end{tabular}
  }
  \caption{Median of $\mathit{CNOT}$ gates, depth, and transpilation time of Grover's algorithm with different number of iterations}
  \vspace*{-6mm}
  \label{table:groverlength}
\end{table}

\subsection{Different Backend Connectivity}
\label{sec:connectivity}


\begin{table}[bhtp]
  \centering
  \small
    \resizebox{\columnwidth}{!}{%
  \begin{tabular}{|c|c|c|c|c|c|c|c|c|}
    \hline
    & \multicolumn{4}{c|}{QPE \texttt{ibmq\_almaden}}
   & \multicolumn{4}{c|}{QPE \texttt{ibmq\_rochester}}\\
     \hline
    Metric & \multicolumn{2}{c|}{$\mathit{CNOT}$ gate count} & \multicolumn{2}{c|}{transpile time(s)}& \multicolumn{2}{c|}{$\mathit{CNOT}$ gate count} & \multicolumn{2}{c|}{transpile time(s)}\\
    \hline
    Optimization & level3 & RPO& level3 & RPO & level3 & RPO & level3 & RPO\\
    \hline
    4-qubits &26 & 23 & 0.24 & 0.23 & 25 & 18 & 0.36 & 0.34\\
    \hline
    6-qubits & 72 & 56 & 0.57 & 0.53 & 66 & 51 & 0.77 & 0.72\\
    \hline
    8-qubits & 157 & 134 & 0.98 & 0.97 & 236 & 220 & 2.45 & 2.42\\
    \hline
    10-qubits & 357 & 312 & 1.62 & 1.54 & 198 & 147 & 1.96 & 1.87\\
    \hline
    12-qubits & 413 & 369 & 2.27 & 2.16 & 444 & 370 & 3.55 & 3.44\\
    \hline
    14-qubits & 586 & 537 & 3.36 & 3.25 & 722 & 644 & 6.17 & 5.77\\
    \hline
  \end{tabular}
  }
  \caption{Median of $\mathit{CNOT}$ gates and transpilation time of QPE algorithm on different quantum computers}
  \label{table:connectivity}
\end{table}

$\mathit{SWAP}$ gates are introduced when the compiler performs the logical-to-physical mapping. When the backend has limited connectivity, the logical-to-physical mapping will introduce more $\mathit{SWAP}$ gates. Therefore, our optimization pass has a higher chance to optimize the quantum circuit.

We compile the QPE program with connectivity maps from three different quantum computers. These connectivity maps are shown in Figure~\ref{fig:connectivity}. Among these quantum computers, \texttt{ibmq\_16\_melbourne} has the best connectivity and \texttt{ibmq\_rochester} has the worst. The experiment result of QPE on \texttt{ibmq\_16\_melbourne} is shown in Table~\ref{table:allbench} in the previous section. The results of QPE on \texttt{ibmq\_almaden} and \texttt{ibmq\_rochester} are shown in Table~\ref{table:connectivity}.\footnote{The abnormal result of 8-qubit QPE algorithm on \texttt{ibmq\_rochester} is
due to Qiskit transpiler's layout selection function changed the shape of the coupling map subgraph}. From these results, we can see that our optimization is effective on all these quantum computers. Another interesting observation is that, the worse connectivity the quantum computer has, the higher total $\mathit{CNOT}$ gate count, and the more $\mathit{CNOT}$ gates will be optimized by our optimization. The percentage of $\mathit{CNOT}$ gates reduced by our optimization are $18.0\%$, $15.2\%$, and $20.6\%$ for \texttt{ibmq\_16\_melbourne}, \texttt{ibmq\_almaden} and \texttt{ibmq\_rochester} respectively. The percentage of transpiled time reduced by our optimization are $5.5\%$, $5.3\%$, and $4.6\%$.

\subsection{Experiment on Real Quantum Computers}

We ran the 3-qubit QPE algorithm on real quantum computers to highlight the effectiveness of reducing $\mathit{CNOT}$ gates. The output distribution is shown in Figure~\ref{fig:QPE_realdevice}. The correct output should be $111$. Since the circuit depth for 3-qubit QPE is shallow, the different quantum computer connectivity doesn't lead to too much difference in the total $\mathit{CNOT}$ count. The gate error and measurement error of different devices have higher impact on the final output distribution. Although the circuit running on \texttt{ibmq\_16\_melbourne} has the least $\mathit{CNOT}$ count, the fidelity of that circuit is not the best. Without our optimization, we cannot even infer the correct result on both \texttt{ibmq\_16\_melbourne} and \texttt{ibmq\_almaden}. Our optimization reduces the $\mathit{CNOT}$ gate count by $33\%$, $29\%$, and $28\%$ and leads to success rate improvements of $2.94X$, $2.69X$, and $1.53X$ on \texttt{ibmq\_16\_melbourne}, \texttt{ibmq\_almaden}, and \texttt{ibmq\_rochester}, respectively. The average success rate improvement (geometric mean) is $2.30X$ for 3-qubit QPE algorithm.

\section{Conclusions}
\label{sec:conclusion}
In this paper, we propose a fast and effective compiler optimization named relaxed peephole optimization. Based on this optimization we designed two compiler optimization passes, QBO and QPO, and implemented them in the IBM's Qiskit transpiler. We show that our optimization pass is faster than the most aggressive optimization level in the Qiskit, and the circuits optimized by our optimization pass also have fewer $\mathit{CNOT}$ gates. Our experiments on the real quantum computers highlight that the reduction in $\mathit{CNOT}$ gate count leads to a significant improvement in the circuit success rate. 

\bibliographystyle{IEEEtranS}
\bibliography{refs}

\appendices


\section{Unitary $U$ and pure state $\ket{\psi}$}
\label{appendix:purestate}
We prove that an n-qubit pure state $\ket{\psi_0}$ can derived by applying an n-qubit unitary gate $U$ to the n-qubit zero state $\ket{0}^{\otimes n}$: $\ket{\psi_0} = U\ket{0}^{\otimes n}$. Based on the state $\ket{\psi_0}$, we can leverage the Gram-Schmidt process~\cite{golub2012gram} to find a set of vectors $\ket{\psi_0}, \ket{\psi_1}, ..., \ket{\psi_{2^n-1}}$ that forms an orthonormal basis. Then, the unitary gate $U$ can be calculated as $U = \ket{\psi_0}\bra{0}^{\otimes n} + \ket{\psi_1}\bra{0}^{\otimes {n-1}}\bra{1} + ... + \ket{\psi_{2^n-1}}\bra{1}^{\otimes n}$.

First, we need to prove $\ket{\psi_0} = U\ket{0}^{\otimes n}$. Since the computational basis is an orthonormal basis, we have $\braket{0|0} = 1$ and $\braket{1|0} = 0$. Therefore,
\begin{multline*}
U\ket{0}^{\otimes n} =  \ket{\psi_0}\braket{0|0}^{\otimes n} + \ket{\psi_1}\braket{0|0}^{\otimes {n-1}}\braket{1|0} + ... \\ ... + \ket{\psi_{2^n-1}}\braket{1|0}^{\otimes n} = \ket{\psi_0}
\end{multline*}

Second, we need to prove the matrix $U$ is a unitary matrix such that we can find a corresponding quantum gate. Since the set of vectors $\{\ket{\psi_i}\}$ form an orthonormal basis, we have $\sum_i \ket{\psi_i}\bra{\psi_i} = I$, here I is the identity matrix. Based on this property, We can prove that $UU^\dagger = \ket{\psi_0}\braket{0|0}^{\otimes n}\bra{\psi_0}$ + $\ket{\psi_1}\braket{0|0}^{\otimes {n-1}}\braket{1|1}\bra{\psi_1}$ + ... + $\ket{\psi_{2^n-1}}\braket{1|1}^{\otimes n}\bra{\psi_{2^n-1}} = \sum_i \ket{\psi_i}\bra{\psi_i} = I$. Similarly, we can prove $U^\dagger U = I$. Therefore, the matrix $U$ is a unitary matrix.

\section{$\mathit{CNOT}$ gate optimization with target qubit in $\ket{-}$ state}
\label{appendix:CNOTminus}
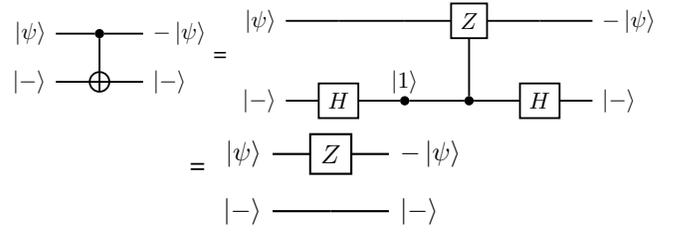
\begin{figure}[!h]
\centering
\begin{adjustbox}{width=\linewidth}
\begin{quantikz}
\lstick{$\ket{\psi}$} &\ctrl{1} & \rstick{$-\ket{\psi}$}\qw \\
\lstick{$\ket{-}$} &\targ{} & \rstick{$\ket{-}$}\qw
\end{quantikz}
=\begin{quantikz}
 \lstick{$\ket{\psi}$}&\qw& \qw & \gate{Z}&\qw & \rstick{$-\ket{\psi}$} \qw \\
 \lstick{$\ket{-}$}& \gate{H} &\phase[label position=above]{\ket{1}}& \ctrl{-1} &\gate{H} & \rstick{$\ket{-}$} \qw
\end{quantikz}
\end{adjustbox}
=\begin{quantikz}
  \lstick{$\ket{\psi}$}& \gate{Z} & \rstick{$-\ket{\psi}$} \qw \\
 \lstick{$\ket{-}$}& \qw & \rstick{$\ket{-}$} \qw
\end{quantikz}
\caption{$\mathit{CNOT}$ gate optimization}
\label{fig:CNOTminus}
\end{figure}

As shown in Figure~\ref{fig:CNOTminus}, the $\mathit{CNOT}$ gate is equivalent to a controlled-$\mathit{Z}$ gate with two Hadamard gates on the target qubit. When the target qubit of the $\mathit{CNOT}$ gate is in $\ket{-}$, after the Hadamard gate, the target qubit is in $\ket{1}$ state. The controlled-$\mathit{Z}$ gate can be optimized into a $\mathit{Z}$ gate, and the two Hadamard gates will be cancelled out. Therefore, the $\mathit{CNOT}$ gate can be substituted with a $\mathit{Z}$ gate on the control qubit.

\section{Closed and open control}
\label{appendix:ClosedandOpen}
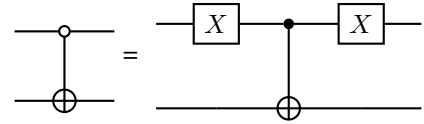
\begin{figure}[ht]
\centering
\begin{center}
\begin{quantikz}
\qw &\octrl{1} & \ghost{X}\qw \\
\qw &\targ{} & \qw
\end{quantikz}
=\begin{quantikz}
\qw & \gate{X} & \ctrl{1} & \gate{X} & \qw \\[2mm]
\qw & \qw      & \targ{}  & \qw      & \qw
\end{quantikz}
\end{center}
\caption{Open and closed control gate equivalence}
\label{fig:openandclosed}
\end{figure}

An open control gate is equivalent to a closed control gate with two $\mathit{NOT}$ gates on the control qubit, as shown in Figure~\ref{fig:openandclosed}. In our paper, we discussed the optimization for the closed control gates. For the open control gates, we can use this equivalence to convert them to the closed control gates and then apply our optimization.

\section{Decomposition of Fredkin gate}
\label{appendix:fredkin}
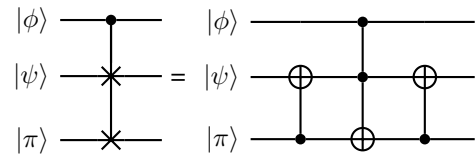
\begin{figure}
\centering
\begin{quantikz}
\lstick{$\ket{\phi}$} &\ctrl{1} & \qw \\
\lstick{$\ket{\psi}$} &\swap{-1} & \qw \\
\lstick{$\ket{\pi}$} &\swap{-1}& \qw
\end{quantikz}
=\begin{quantikz}
\lstick{$\ket{\phi}$} & \qw & \ctrl{1} & \qw & \qw\\
\lstick{$\ket{\psi}$} & \targ{} &\ctrl{1} & \targ{} & \qw \\
\lstick{$\ket{\pi}$} &\ctrl{-1} & \targ{} &\ctrl{-1} & \qw
\end{quantikz}
\caption{Decomposition of Fredkin gate}
\label{fig:fredkin}
\end{figure}

The decomposition of Fredkin gate is shown in Figure~\ref{fig:fredkin}

\section{Additional Experiment Results}
\label{appendix:additional results}

The detailed experiment results for QPE, VQE, quantum volume, and Grover's algorithms are included in Table~\ref{table:depthsinglegate}.

\begin{table*}[htbp]
  \centering
  \small
    \resizebox{\linewidth}{!}{%
  \begin{tabular}{|c|c|c|c|c|c|c|c|c|c|c|c|c|c|c|c|c|c|c|c|c|c|c|c|c|}
    \hline
    & \multicolumn{6}{c|}{QPE}
   & \multicolumn{6}{c|}{VQE}
   &\multicolumn{6}{c|}{Quantum Volume} 
   &\multicolumn{6}{c|}{Grover} \\
     \hline
    Metric & \multicolumn{3}{c|}{single-qubit gate count} &  \multicolumn{3}{c|}{depth}& \multicolumn{3}{c|}{single-qubit gate count} & \multicolumn{3}{c|}{depth}& \multicolumn{3}{c|}{single-qubit gate count} & \multicolumn{3}{c|}{depth}& \multicolumn{3}{c|}{single-qubit gate count} & \multicolumn{3}{c|}{depth}\\
    \hline
    Optimization & level3 & hoare & RPO& level3 & hoare & RPO & level3 & hoare & RPO & level3 & hoare & RPO & level3 & hoare & RPO & level3 & hoare & RPO& level3 & hoare & RPO & level3 & hoare & RPO\\
    \hline
    4-qubits & 65 & 65 & 59 & 45 & 43 & 38 & 136 & 133 & 129 & 97 & 93 & 89 & 107 & 107 & 103 & 44 & 43 & 43 & 375 & 375 & 368 & 329 & 324 & 318 \\
    \hline
    6-qubits & 171 & 169 & 154 & 93 & 92 & 87 & 338 & 340 & 327 & 186 & 185 & 181 & 275 & 275 & 270 & 108 & 108 & 103 & 834 & 834 & 798 & 720 & 712 & 658\\
    \hline
    8-qubits & 351 & 344 & 327 & 149 & 145 & 134 & 686 & 663 & 647 & 310 & 307 & 300& 500 & 498 & 479 & 500 & 497 & 493 & 3666 & 3666 & 3502 & 2963 & 2955 & 2874\\
    \hline
    10-qubits & 621 & 621 & 591 & 210 & 208 & 196 & 1061 & 1056 & 1053 & 413 & 410 & 406 & 802 & 802 & 758 & 211 & 209 & 193 & 14674 & 14666 & 14455 & 12219 & 12181 & 12012 \\
    \hline
    12-qubits & 879 & 877 & 832 & 259 & 258 & 243 & 1591 & 1582 & 1570 & 534 & 532 & 526 & 1199 & 1199 & 1163 & 261 & 258 & 244 & 58424 & 58242 & 57874 & 48315 & 48219 & 47899\\
    \hline
    14-qubits & 1211 & 1211 & 1161 & 389 & 389 & 368 & 2268 & 2340 & 2263 & 751 & 751 & 751 & 2903 & 2910 & 2889 & 946 & 951 & 931 & 232290 & N.A. & 241716 & 192526 & N.A. & 192487\\
    \hline
    
  \end{tabular}
  }
  \caption{Median of single-qubit gates and circuit depth of three quantum algorithms with different size (on \texttt{ibmq\_16\_melbourne})}
  \label{table:depthsinglegate}
\end{table*}

\section{Swap gate on basis-state}
\label{appendix:swap}
Table~\ref{tbl:SWAP} shows QBO implementation on the $\mathit{SWAP}$ gate. 
Take as an example the case where the $\mathit{SWAP}$ gate has the top input in the $\ket{0}$ state and the bottom input in the $\ket{-}$ state.
Since $\ket{0} = X\ket{1}$ and $\ket{1} = h\ket{-}$, the $\ket{0}$ state can be obtained by applying $\mathit{H}$ and $\mathit{X}$ gate to the $\ket{-}$ state, $\ket{0} = X\ket{1} = XH\ket{-}$. Similarly, the $\ket{-}$ can be obtained by applying $\mathit{X}$ and $\mathit{H}$ gate to the $\ket{0}$ state.

\begin{table}[h]
\centering
\caption[]{Equivalences for basis-states in 
\begin{adjustbox}{height=\baselineskip}
\begin{quantikz}[row sep=3mm, column sep=3mm]
\lstick{$\ket{\phi}$} &\swap{1} & \qw \\
\lstick{$\ket{\psi}$} &\swap{-1} & \qw
\end{quantikz}
\end{adjustbox}
}
\label{tbl:SWAP}
\begin{adjustbox}{width=\linewidth, height = 3.2cm}
$\begin{array}{r|c|c|c|c|c}
\tikz{\node[below left, inner sep=1pt] (target) {$ \ket{\psi} $};%
      \node[above right,inner sep=1pt] (control) {$\ket{\phi}$};%
      \draw (target.north west|-control.north west) -- (target.south east-|control.south east);}
& 
\multicolumn{1}{c}{\top} & 
\multicolumn{1}{c}{\ket{0}} &
\multicolumn{1}{c}{\ket{1}} &
\multicolumn{1}{c}{\ket{+}} & 
\multicolumn{1}{c}{\ket{-}} \\
\cline{1-6} 
\top
&
\begin{tikzcd}
\qw & \swap{1}  & \qw \\ 
\qw & \swap{-1} & \qw    
\end{tikzcd}
&
\begin{tikzcd}
\qw & \aswap & \qw \\  
\qw & \swap{-1} & \qw  
\end{tikzcd}&
\begin{tikzcd}
\qw & \qw & \aswap & \qw \\       
\qw & \gate{X} & \swap{-1} & \qw  
\end{tikzcd}&
\begin{tikzcd}
\qw & \swap{1} & \qw \\
\qw & \aswap & \qw
\end{tikzcd}&
\begin{tikzcd}
\qw & \qw & \swap{1} & \qw \\
\qw & \gate{Z} &\aswap  & \qw
\end{tikzcd}\\
\cline{2-6} 
\ket{0} &
\begin{tikzcd}
\qw &\swap{1} & \qw \\ 
\qw &\aswap & \qw     
\end{tikzcd} &
\begin{tikzcd}
\qw & \qw & \qw \\[2mm]  
\qw & \qw & \qw            
\end{tikzcd} &
\begin{tikzcd}
\qw & \gate{X} & \qw \\     
\qw & \gate{X} & \qw   
\end{tikzcd} & 
\begin{tikzcd}
\qw & \gate{H} & \qw \\    
\qw & \gate{H} & \qw       
\end{tikzcd} & 
\begin{tikzcd}
\qw & \gate{H} & \gate{X}  & \qw \\    
\qw & \gate{X} & \gate{H} \qw       
\end{tikzcd} \\
\cline{2-6} 
\ket{1} &
\begin{tikzcd}
\qw & \gate{X} & \swap{1} & \qw \\
\qw & \qw & \aswap & \qw
\end{tikzcd} &
\begin{tikzcd}
\qw & \gate{X} & \qw \\  
\qw & \gate{X} & \qw            
\end{tikzcd}&
\begin{tikzcd}
\qw & \qw & \qw \\[2mm]     
\qw & \qw & \qw   
\end{tikzcd} 
&
\begin{tikzcd}
\qw & \gate{H} & \gate{X} & \qw \\ 
\qw & \gate{X} & \gate{H} & \qw   
\end{tikzcd} 
& 
\begin{tikzcd}
\qw & \gate{H} & \qw \\ 
\qw & \gate{H} & \qw   
\end{tikzcd} 
\\
\cline{2-6} 
\ket{+} &
\begin{tikzcd}
\qw &\aswap & \qw \\      
\qw &\swap{-1} & \qw           
\end{tikzcd} &
\begin{tikzcd}
\qw & \gate{H} & \qw \\      
\qw & \gate{H} & \qw           
\end{tikzcd}
&
\begin{tikzcd}
\qw & \gate{X} & \gate{H} & \qw \\
\qw & \gate{H} & \gate{X} & \qw
\end{tikzcd}
&
\begin{tikzcd}
\qw & \qw & \qw \\[2mm]
\qw & \qw & \qw
\end{tikzcd}
&
\begin{tikzcd}
\qw &\gate{Z} & \qw \\
\qw &\gate{Z} & \qw
\end{tikzcd}
\\
\cline{2-6} 
\ket{-} &
\begin{tikzcd}
\qw & \gate{Z} & \aswap & \qw \\
\qw & \qw & \swap{-1} & \qw
\end{tikzcd}
&
\begin{tikzcd}
\qw & \gate{X} & \gate{H} & \qw \\
\qw & \gate{H} & \gate{X} & \qw
\end{tikzcd}
&
\begin{tikzcd}
\qw & \gate{H} & \qw \\
\qw & \gate{H} & \qw
\end{tikzcd}
&
\begin{tikzcd}
\qw & \gate{Z} & \qw \\ 
\qw & \gate{Z} & \qw    
\end{tikzcd}
&
\begin{tikzcd}
\qw & \qw & \qw \\[2mm] 
\qw & \qw & \qw         
\end{tikzcd}
\\
\end{array}$
\end{adjustbox}
\end{table}

\newpage
\section{Artifact Description Appendix}

\subsection{Abstract}

Our artifact provides the experiments for all our evaluated benchmarks, along with the experiments to validate the quantum circuit costs.

We also provide the source code for our compiler optimization passes and all of our benchmarks.

\subsection{Artifact check-list (meta-information)}

\small
\begin{itemize}
  \item {\bf Algorithm: }Peephole optimization algorithm
  \item {\bf Data set: } Benchmarks included in our paper
  \item {\bf Hardware: }We recommend running the experiments on a 15-qubit IBM Q machine to verify the results
  \item {\bf Execution: }Run the corresponding python scripts and jupyter notebooks
  \item {\bf Metrics: }
  CNOT gates: The number of CNOT gates in the quantum program.
  
  transpile time: The time for quantum circuit transpilation.
  
  single-qubit gate count: The number of single-qubit gates in the quantum circuit.
  
  depth: The depth of the quantum circuit.
  
Success rate: The ratio of the count of correct output state over the total count of the trials.
  \item {\bf Output: } The CNOT gates, transpilation time, single-qubit gate count, and depth will be dumped in a csv file in the results folder. The output distribution of the experiments will be printed in the corresponding jupyter notebooks.
  \item {\bf Experiments: } We use functions from Qiskit to calculate the number of gates and transpilation time of our circuit. We calculate the success rate based on the output distribution from the IBMQ backends.
  \item {\bf How much disk space required (approximately)?: }2GB
  \item {\bf How much time is needed to prepare workflow (approximately)?: }A couple of minutes.
  \item {\bf How much time is needed to complete experiments (approximately)?: } ~18 hours in total.
  \item {\bf Publicly available?: } Yes.
  \item {\bf Code licenses (if publicly available)?: } Apache-2.0 License
  \item {\bf Archived (provide DOI)?: } \href{https://zenodo.org/record/4281275#.X7fI72hKg2w}{10.5281/zenodo.4281275}
\end{itemize}

\subsection{Description}
\subsubsection{How delivered}
Our source code, benchmarks, and jupyter notebooks for experiments are available on Github: \href{https://github.com/1ucian0/rpo.git}{https://github.com/1ucian0/rpo.git}

\subsubsection{Hardware dependencies}
In our paper, we run our experiments on 15-qubit quantum computer ibmq\_16\_melbourne, a 20-qubit quantum computer ibmq\_almaden, and a 53-qubit quantum computer ibmq\_rochester. Since some of the quantum computers are not publicly available, in order to reproduce the results, we use the fakebackends from Qiskit to use the actual device configurations, such as coupling maps. 
\subsubsection{Software dependencies}
Python version $\geq3.5, <3.9$, Qiskit 0.18.0, Jupyter notebook, matplotlib 3.3, z3-solver, tabulate. 

Qiskit requires Ubuntu 16.04 or later, MacOS 10.12.6 or later, or Windows 7 or later.
\subsubsection{Data sets}
Quantum computing benchmarks mentioned in our paper.
\subsection{Installation}
We recommend installing the software in an Anaconda environment with Python version 3.7.
After downloading Anaconda, create an environment:

\begin{verbatim}
$ conda create -n my_env python=3.7
\end{verbatim}

Then, activate the environment:
\begin{itemize}
  \item For Linux or MacOS: \texttt{\$ source activate my\_env} 
  \item For Windows: 
  \texttt{\$ activate my\_env}
\end{itemize}

You can clone our source code and benchmarks from GitHub:

\begin{verbatim}
$ git clone https://github.com/1ucian0/rpo.git
\end{verbatim}

After cloning the GitHub repository, to install the required software:

\begin{verbatim}
$ pip install -r requirements.txt
\end{verbatim}

For questions regarding Qiskit installation, please refer to: 

\href{https://qiskit.org/documentation/install.html }{https://qiskit.org/documentation/install.html }

After installation, run the unittests to verify the installation:

\begin{verbatim}
$ python -m unittest discover -v tests
\end{verbatim}
\subsection{Experiment workflow}
The experiment results on transpilations (Table II, III, IV, V) can be verified by running the \texttt{run\_benchmark.py} file with corresponding arguments. For example, the following command is for running the Quantum Phase Estimation (QPE) benchmark on ibmq\_16\_melbourne backend:

\begin{lstlisting}[breaklines, basicstyle=\ttfamily, numbers=none,]
$ python run_benchmark.py benchmark/qpe_FakeMelbourne.yaml
\end{lstlisting}

The results will be dumped in \texttt{results/qpe\_FakeMelbourne.csv} in this case. In general, \texttt{python run\_benchmark.py benchmark/some\-thing.yaml} dumps its result in \texttt{results/something.csv}.

The CSV files contain all the raw data of multiple runs. Execute the \texttt{table.py} file to generate the table containing medians of multiple runs:

\begin{verbatim}
$ python table.py  results/qpe_FakeMelbourne.csv
\end{verbatim}

To verify the experiments on the real quantum computers (Figure.11), run the corresponding Jupyter notebooks: \texttt{QPE\_almaden/melbourne/rochester}. The results will be printed in the Jupyter notebook output.

\subsection{Evaluation and expected result}

The results are dumped after running the corresponding Python script and Jupyter notebook. The Jupyter notebooks contain prior experimental results reported in our paper. The expected results are listed in our paper (Table II,III,IV,V, Figure.11).

\subsection{Experiment customization}
The YAML files and Jupyter notebooks are all customizable to run different benchmarks with different configurations. The parameters of the YAML files are described in \texttt{README.md} file.
\subsection{Notes}
Some of the experiments (QPE\_almaden.ipynb, QPE\_rochester.ipynb) require hardware access to certain IBMQ quantum computers. The user can use publically available machines to verify the result.

The transpilation time for Grover's algorithm is particularly long, to validate the results, we changed the number of experiments to five times.
\subsection{Methodology}

Submission, reviewing and badging methodology:

\begin{itemize}
  \item \url{http://cTuning.org/ae/submission-20190109.html}
  \item \url{http://cTuning.org/ae/reviewing-20190109.html}
  \item \url{https://www.acm.org/publications/policies/artifact-review-badging}
\end{itemize}

\end{document}